\documentclass[twocolumn,aps,preprintnumbers,amsmath,amssymb,nofootinbib,floatfix,pre]{revtex4-1}
\usepackage[dvips]{graphicx}
\usepackage{graphicx}
\usepackage{amssymb,amsmath}

\begin{document}

\newcommand{\BejLen}{l_{\mathrm{B}}}
\newcommand{\AArm}{\, \mathrm{\AA}}
\newcommand{\Mrm}{\, \mathrm{M}}
\newcommand{\rbf}{\mathbf{r}}
\newcommand{\kbf}{\mathbf{k}}
\newcommand{\Gk}{\widetilde{G}(\mathbf{k})}
\newcommand{\ehat}{\mathbf{\hat{e}}}
\newcommand{\KBoltz}{k_{\mathrm{B}}}
\newcommand{\DebLen}{\lambda_{\mathrm{D}}}
\newcommand{\DebFreq}{\kappa_{\mathrm{D}}}
\newcommand{\captionset}{\addtolength{\leftskip}{.3cm}\addtolength{\rightskip}{.4cm} \small\textsf}

\title{Dipolar Poisson-Boltzmann Approach to Ionic Solutions: \\A Mean Field and Loop Expansion Analysis}

\author{Amir Levy, David Andelman$^{*}$}
\affiliation{Raymond and Beverly Sackler School of Physics and Astronomy,
Tel Aviv University, Ramat Aviv 69978, Tel Aviv, Israel}

\author{Henri Orland}
\affiliation{Institut de Physique Th\'eorique, CE-Saclay, CEA,
F-91191 Gif-sur-Yvette Cedex, France}

\date{11.8.2013}

\begin{abstract}

We study the variation of the dielectric response of ionic aqueous solutions as function of their ionic strength.
The effect of salt on the dielectric constant appears through the coupling between
ions and dipolar water molecules.
On a mean-field level, we account for any internal charge distribution of particles. The dipolar degrees of freedom are added to the ionic ones and result in a generalization of the Poisson-Boltzmann (PB) equation called the Dipolar PB (DPB).
By looking at the DPB equation around a fixed point-like ion, a closed-form formula for the dielectric constant is obtained. We express the dielectric constant using  the ``hydration length" that characterizes the hydration shell of dipoles around ions, and thus the strength of the dielectric decrement. The DPB equation is then examined for three additional cases: mixture of solvents, polarizable medium and  ions of finite size.
Employing field-theoretical methods we expand the Gibbs free-energy to first order in
a loop expansion and calculate self-consistently the dielectric constant.
For pure water, the dipolar fluctuations represent an important correction to
the mean-field value and good agreement with the water dielectric constant is obtained. For ionic solutions
we predict analytically the dielectric decrement that depends on the ionic
strength in a non-linear way. Our prediction fits rather well a
large range of concentrations for different salts using only one fit parameter related to
the size of ions and dipoles.
A linear dependence of the dielectric constant on
the salt concentration is observed at low salinity, and a noticeable deviation
from linearity can be seen for ionic strength above 1\,M, in agreement with experiments.

\end{abstract}

\maketitle

\section{Introduction}

The electrostatic interactions between charges in aqueous solutions play an important role in chemistry, biology and materials science. The Poisson-Boltzmann (PB) theory gives a simple yet powerful description for such systems, taking into account only the Coulombic forces on a mean-field level \cite{andelman1, Israelachvili}. Despite its limitations, the PB theory succeeds in capturing the main features of the underlying physics for monovalent ions and weak surface charges.

Since the PB theory is a mean-field approximation, it does not take into account neither the correlations between the charges, nor does it allow for fluctuations around the mean-field solution, and over the years several alternatives and extensions of this theory have been proposed.  They include significant corrections in cases of multivalent ions and high charge density, especially near surfaces and membranes, and the effects of correlations and fluctuations \cite{DPB,ByondPB,MoreiraNetz,IntegralEqn1,IntegralEqn2}. For very high ionic densities, steric effects prevent ions from accumulating near charged surfaces, and lead to a modified PB (MPB) equation
\cite{Steric1,Steric2,IonSpecInt}.
Other interactions such as van der Waals can be added to the electrostatic ones, resulting in the well known DLVO (Deryagin-Landau-Verwey-Overbeek) theory \cite{DLVO},
which successfully explains stability of charged colloidal suspensions.
More recently, molecular dynamics (MD) simulations have been used to study the behavior of aqueous solutions, allowing the study of very specific models for  solvent and solute molecules~\cite{MonteCarlo, MolecularDynamics1,MD1, MD2, MD3}.

Another impediment of PB theory is that it fails to account for the dielectric constant decrement of ionic solutions.
The overall change in the dielectric constant of an ionic solution can be large, and lead to significant differences in the behavior of ionic solutions near interfaces and surfaces and to other ion-specific effects~\cite{AqueousDielectrics, PolarMolecules,Hasted,IonSpecific, IonSpec2, IonSpec3}.

\begin{figure}[ht!]
  \centering
    \includegraphics[width=0.4\textwidth]{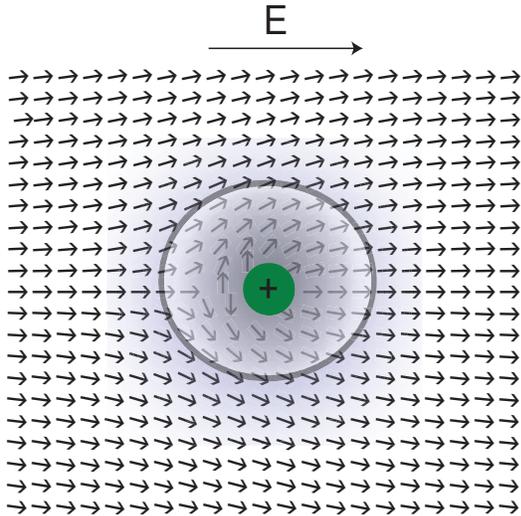}
\caption{\captionset{(color online) A schematic drawing of the dipolar response to a central charge following Eq.~(\ref{D0Solution}). The arrows are aligned along the direction of the local electric field created by a positive charge placed at the origin, as well as by the constant external field $\mathbf{E}$. The hydration shell, which is the area most effected by the charged particle, is encircled. }}
\label{fig1}
\end{figure}

The ions affect the dielectric constant via two principal mechanisms. The first is the polarizability of the ions themselves~\cite{ElecChemDep}. The second and more important  is due to the $\emph{hydration shell}$~\cite{IonSpecific,Gluekauf} as shown in Fig.~\ref{fig1}. The hydration shell is created by the interactions between the molecules of the dielectric medium (water) and the ions. The strong electric field around each ion is greater than the external electric field, and re-orients the dipoles in its vicinity. The total response of dipoles to the external field is thus smaller and leads to a reduction in the dielectric  constant.

Both of these mechanisms, at least for dilute solutions, are linear in the ionic concentration. As long as the hydration shell radius is smaller than the distance between neighboring ions, each ion contributes for the decrement of the total dielectric constant independently of the other ions. This linear dependence of the dielectric constant on the concentration can be written as:
\begin{eqnarray}
\label{DecrementForm}
\varepsilon(n_s) = \varepsilon_w + \gamma n_s ,
\end{eqnarray}
where $\varepsilon_w$ is the pure water dielectric constant, $n_s$ is the ionic (salt) concentration and $\gamma$ is the linear term coefficient. The value of $\gamma/\varepsilon_0$ ($\varepsilon_0$ is the vacuum permittivity) is ion dependent~\cite{Experiment1,Experiment2,Experiment3} and ranges from $-8 \mathrm{M}^{-1}$ to $-20 \mathrm{M}^{-1}$, for ionic concentrations up to $1$\,M.

In this paper we go beyond the basic PB theory, extending and elaborating on our recent Letter~\cite{PRL}. Three major modifications of the PB theory are considered: first we relax the assumption of the continuous water dielectric medium and consider instead a microscopic model of dipoles. Second, we take into account fluctuations and correlations between dipoles and ions via a field-theoretical loop expansion. Finally, we allow the charges to have a finite size and also consider mixtures of dipoles and the case of polarizable dipoles. Other phenomena such as non-Coulombic interactions and dynamical effects~\cite{MolecularDynamics1, MolDynDec} will not be taken into account, in order to keep the model as simple as possible.

The outline of our paper is as follows. We begin in Sec.~II by reconstructing a generalized PB theory from a grand-canonical ensemble of charged particles with arbitrary internal charge distribution. We then focus in Sec.~III on the specific Dipolar Poisson-Boltzmann (DPB) equation and extract the dielectric constant for several interesting cases, following (Sec.~IV) by a loop-expansion calculation for the influence of correlations. To first order beyond mean field, a closed formula for the dielectric constant is obtained and we show its agreement with experimental data in Sec.~V. Finally, in Sec.~VI we conclude with some remarks and future prospects.

\section {The Model}

We consider a system with several types of charged particles. The particles can be dipoles, counter-ions, etc. Each type of particle is characterized by its internal charge distribution. A fixed (``frozen") charged distribution that can represents fixed surface charges (or other boundary) is also included.
The total charge density of a mixture of different charged particles can be written as:
\begin{eqnarray}
\label{chargdist}
\rho(\rbf) & = & \sum_{l=1}^{M}\sum_{i=1}^{N_l}\rho_{l}(\Omega_{il};\rbf-\rbf_{il}) + \rho_f(\rbf),
\end{eqnarray}
where $M$ is the number of different types of particles, $N_l = N_1,N_2,\ldots,N_M$ is the number of particles of the $l^{th}$ type, $\rho_{l}(\Omega_{il};\rbf-\rbf_{il})$ is the charge density profile of the $i^{th}$ particle of the $l^{th}$ type rotated by a spatial (solid) angle $\Omega_{il}$ and located at position $\rbf_{il}$. The spatial angle $\Omega_{il}$ is composed of an azimuth angle $\phi_{il}$ and an elevation angle $\theta_{il}$, so all possible rotations are accounted for. Charges of the same type have the same charge distribution, up to changes in their location and orientation. Finally, $\rho_f(\rbf)$ is an added fixed charge distribution.

Assuming Coulombic interactions between any two charges, the grand-canonical partition function can be written as:
\begin{eqnarray}
\label{Zdef}
\Xi  & = & \sum_{N_1=1}^{\infty}\frac{(\Lambda_1)^{N_1}}{N_1!}  \sum_{N_2=1}^{\infty} \frac{(\Lambda_2)^{N_2}}{N_2!}\ldots
\sum_{N_M=1}^{\infty}\frac{(\Lambda_M)^{N_M}}{N_M!}
\nonumber\\
& \times &  \int\prod_{l=1}^M\prod_{i=1}^{N_l}{\rm d}^3\rbf_{il}\frac{{\rm d}^2 \Omega_{il}}{4\pi}
\nonumber\\
& \times & \exp \left[-\frac{\beta}{2} \int \mathrm{d}^3\rbf \, \mathrm{d}^3 \mathbf{r'} \rho(\rbf)v(\mathbf{r-r'})\rho(\mathbf{r'})\right]\, ,
\end{eqnarray}
where $v(\mathbf{r-r'}) = 1/ (4\pi\varepsilon_0 |\rbf - \rbf'|)$ is the Coulomb potential between any two unit charges, $\beta = 1/\KBoltz T$ is the inverse thermal energy, $\Lambda_l = \exp(\beta \mu_l)$ is the fugacity for the $l^{th}$ particle type, and $\mu_l$ is their chemical potential. We employ the Hubbard-Stratonovich transformation~\cite{HubStrato}, which introduces a new auxiliary field, $\phi(\rbf)$, coupled with the charge density $\rho(\rbf)$:
\begin{eqnarray}
\label{expbetaH}
& & \exp \left[-\frac{\beta}{2} \int \mathrm{d}^3\rbf \, \mathrm{d}^3 \mathbf{r'} \rho(\rbf)v(\mathbf{r-r'})\rho(\mathbf{r'})\right]
\nonumber\\
& = & \int{\cal D} \phi(\rbf)\exp \left[-\frac{\beta}{2 } \int {\rm d}^3\rbf \,{\rm d}^3 \mathbf{r'}\phi(\rbf) v^{-1}(\rbf-\mathbf{r'})\phi(\mathbf{r'})\right.
\nonumber\\
&&\left. -i \beta \int {\rm d}^3\rbf \, \phi(\rbf)\rho(\rbf)\right] \,.
\end{eqnarray}
The inverse Coulomb potential is equal to $v^{-1}=-\varepsilon_0 \nabla^2\delta(\rbf-\mathbf{r'})$, as can be seen from  Poisson equation. The Hubbard-Stratonovich transformation is used to linearize the interaction term in the partition function, Eq.~(\ref{Zdef}). By combining the general charge distribution, Eq.~(\ref{chargdist}), the partition function reads:
\begin{eqnarray}
\label{GC}
\Xi & = &  \int{\cal D} \phi(\rbf) \,  \mathrm{e}^{- \beta F[\phi(\rbf)]},
\end{eqnarray}
where the free energy functional $F$ is defined as:
\begin{eqnarray}
\label{GC_F}
- \beta F & = & -\frac{\varepsilon_0\beta}{2}\int {\rm d}^3\rbf \, [\nabla\phi(\rbf)]^2
\nonumber\\
& -& i\beta \int {\rm d}^3 \rbf \, \phi(\rbf) \rho_f(\rbf) + \sum_{l=1}^M \Lambda_l \int {\rm d}^3 \rbf\int \frac{{\rm d}^2 \Omega}{4\pi}
\nonumber\\
& \times &  \, \exp \left[ -i \beta \int {\rm d}^3 \rbf' \, \rho_l(\Omega; \rbf'-\rbf) \phi(\mathbf{r'})\right].
\end{eqnarray}
The partition function in Eq.~(\ref{GC}) has the form of a functional integral over all possible configurations of $\{\phi(\rbf)\}$.
The electrostatic potential $\psi$ is derived from the grand-canonical partition function by adding a ghost source term, $\rho_0$:
\begin{eqnarray}
\label{eqghost}
\rho(\rbf) & \rightarrow & \rho(\rbf) + \rho_0(\rbf),
\nonumber\\
\psi(\rbf) & = & -\frac{1}{\beta} \left. \frac{\delta \ln \Xi[\rho_0(\rbf)]}{\delta \rho_0(\rbf)}\right|_{\rho_0=0}\, .
\end{eqnarray}
From Eq.~(\ref{GC_F}) we can see that adding a  fixed charge distribution to the charge density as in Eq.~(\ref{eqghost}), will add to the grand-canonical partition function the following term:
\begin{eqnarray}
\Xi [\rho_0(\rbf)] =  \int{\cal D} \phi(\rbf) \exp \left(- \beta F  -  i \beta \int {\rm d}^3 \rbf \, \phi(\rbf) \rho_0(\rbf) \right).
\,\,\,\,\,\,\,\,
\end{eqnarray}
Let us denote the grand-canonical partition function with no added source term ($\rho_0 = 0$) as $\Xi_0$:
\begin{eqnarray}
\label{XsiPathIntDef}
\Xi_0 &=&  \int{\cal D}\phi(\rbf) \exp \left(- \beta F \right).
\end{eqnarray}
The electrostatic potential $\psi$ is equal to:
\begin{eqnarray}
\psi &=& \frac{i}{\Xi_0} \int{\cal D} \phi(\rbf)\,\, \phi(\rbf) \exp \left(-\beta  F  \right) = \langle i \phi \rangle,
\end{eqnarray}
where $\langle \ldots \rangle$ denotes thermodynamical averaging. In the mean-field approximation only the saddle point of the action contributes to the functional integral, and the electrostatic field exactly equals to $i \phi$, $\psi=i\phi$. Similarly, it can be shown \cite{ColFluid} that on the mean-field level the fugacity of any charge type equals to it corresponding bulk charge density, $\Lambda_l = n_l$.

\subsection{PB equation and the Debye-H\"uckel approximation}

The PB equation for the electrostatic potential can be derived as the saddle point of the grand-canonical partition function. Writing the Euler-Lagrange equation for the functional $F$ yields an equation for the mean-field value of the auxiliary field $\phi(\rbf)$. On the same level of approximation, as noted before,  $\phi(\rbf)=-i\psi(\rbf)$, and $\Lambda_l=n_l$, and the Euler-Lagrange equation is an integro-differential equation that constitutes a generalization of the PB equation:
\begin{eqnarray}
\label{expfullpb}
-\varepsilon_0 \nabla^2\psi(\rbf) & = &  \rho_f(\rbf)+ \int \frac{{\rm d}^2 \Omega}{4\pi} \sum_{l=1}^M \Lambda_l \int {\rm d}^3 \mathbf{r''} \, \rho_{l}(\Omega; \rbf-\rbf'')
\nonumber\\
& \times & \exp \left[ -\beta \int {\rm d}^3 \rbf' \, \rho_l(\Omega; \rbf'-\rbf'') \psi(\rbf')\right] \, ,
\end{eqnarray}
where the standard PB form is recovered for point-like charges with charge $q_l$: $\rho_l=q_l\sum_{i=1}^{N_l}\delta(\rbf-\rbf_{il})$.

An alternative and more compact way of writing Eq.~(\ref{expfullpb}) is:
\begin{eqnarray}
& & -\varepsilon_0 \nabla^2\psi(\rbf)    =  \rho_f(\rbf)
\nonumber\\
& + & \Big\langle \sum_{l=1}^M \Lambda_l \rho_{l}(\Omega; -\rbf) \otimes \exp\left[ -\beta \rho_{l}(\Omega;\rbf) \otimes \psi(\rbf)\right]\Big\rangle_\Omega,
\,\,\,\,\,\,\,
\end{eqnarray}
where $\langle \ldots \rangle_\Omega$ denotes orientation averaging and $\otimes$ stands for the convolution operation:
\begin{eqnarray}
f(\rbf) \otimes  g(\rbf) \equiv \int {\rm d}^3 \mathbf{r'}\, f(\mathbf{r-r'}) g(\mathbf{r'}).
\end{eqnarray}
Replacing the point-like particles in the original PB model with charged particles having a more complicated internal charge distribution is at the origin of the non-locality of the above Eq.~(\ref{expfullpb}).

In the Debye-H\"uckel (DH) approximation an exact solution can be derived, in a way that would illustrate the size effect of charged particles. The linear DH equation is calculated by expanding the exponent in Eq.~(\ref{expfullpb}) to first order:
\begin{eqnarray}
\label{expfullpb_linear}
& & -\varepsilon_0 \nabla^2\psi(\rbf)  =  \sum_{l=1}^M n_l q_l + \rho_f(\rbf)
\nonumber\\
& -&  \beta \sum_{l=1}^M n_l \Big\langle\rho_{l}(\Omega;\mathbf{-r}) \otimes\rho_{l}(\Omega;\rbf) \otimes \psi(\rbf) \Big\rangle_\Omega,
\end{eqnarray}
where $n_l$ is the bulk value of the number charge density of the $l^{th}$  particle type, and $q_l$ is the total charge $q_l = \int {\rm d}^3 \rbf\, \rho_l(\rbf)$. The first term in Eq.~(\ref{expfullpb_linear}) can be omitted because of electro-neutrality, $ \sum_{l=1}^M n_l q_l =0$. Denoting $\tilde{\psi}(\kbf)$, $\tilde{\rho_f}(\kbf)$ and $\tilde{\rho}_l(\kbf)$ as the Fourier transform of $\psi(\rbf)$, $\rho_f(\rbf)$ and $\rho_l(\rbf)$, respectively, the PB equation takes the following form:
\begin{eqnarray}
\label{dheqn}
\varepsilon_0 k^2\tilde{\psi}(\kbf)  =
 \tilde{\rho_f}(\kbf) - \beta\sum_{l=1}^M n_l\langle |\tilde{\rho_l}(\kbf)|^2\rangle_\Omega\tilde{\psi}(\kbf) ,
\end{eqnarray}
where $\tilde{f}(\kbf)=\int {\rm d}^3 \rbf f(\rbf) {\rm e}^{i\kbf \cdot \rbf}$ is the Fourier transform of $f(\rbf)$. In comparison with the Fourier transform of the standard DH equation, the key difference is that the net charge term, $q_l^2$, is replaced  by the charge structure factor, $S_l(k)=\left\langle|\tilde{\rho_l}(\kbf)|^2\right\rangle_\Omega$. This difference is observable only for length scale comparable with the size of the particles. In the large distance limit, $r\rightarrow \infty$, corresponding to short wavenumbers, $k \rightarrow 0$:
\begin{eqnarray}
\left.\tilde{\rho}_l(\kbf)\right|_{k= 0} = \int {\rm d}^3 \rbf\, \rho_l(\rbf) = q_l,
\end{eqnarray}
and thus the generalized DH equation goes back to the regular DH one.

The general solution for the DH equation can be written in an integral form for the electrostatic potential $\psi$:
\begin{eqnarray}
\label{psikeqn}
\psi(\rbf) = \int \frac{{\rm d}^3 \kbf}{(2\pi)^{3}} \frac{\tilde{\rho}_f(\kbf)}{\varepsilon_0k^2+\beta\sum_{l=1}^M n_l S_l(k)}{\rm e}^{-i \kbf \cdot \rbf}.
\end{eqnarray}
As we can see from Eq.~(\ref{psikeqn}), the vacuum permittivity $\varepsilon_0$ is the coefficient of $k^2$ in the denominator, while the combined coefficient of all the $k^2$ terms contributes to the effective overall dielectric constant, $\varepsilon$. Expanding $S_l(k)$ in Taylor series up to $k^2$:
\begin{eqnarray}
\label{Sk2}
S_l(k) &\approx & \left( \int \rho_l(\rbf)\, {\rm d}^3 \rbf\right)^2+ \frac{k^2}{3} \left( \int {\rm d}^3 \rbf\, \rho_l(\rbf) \rbf \right)^2
\nonumber\\
& - & \frac{k^2}{3} \int {\rm d}^3 \rbf\, \rho_l(\rbf) \int {\rm d}^3 \rbf\, \rho_l(\rbf) \rbf^2,
\end{eqnarray}
and substituting Eq.~(\ref{Sk2}) into  Eq.~(\ref{psikeqn}), gives us a closed-form formula for the medium overall dielectric constant, $\varepsilon$:
\begin{eqnarray}
\label{epsExpansion}
\varepsilon = \varepsilon_0 + \frac{\beta}{3} \sum_{l=1}^M n_l p_l^2 -  \frac{\beta}{3}\sum_{l=1}^M  n_l q_l \int {\rm d}^3 \rbf\, \rho_l(\rbf)r^2,
\end{eqnarray}
where ${\bf p}_l$ is the dipole moment of the $l^{th}$ type:
\begin{eqnarray}
{\bf p}_l = \int {\rm d}^3 \rbf\, \rho_l(\rbf) \rbf.
\end{eqnarray}

The second term in Eq.~(\ref{epsExpansion}) has exactly the same form as for point-like dipoles~\cite{DPB}, but is derived here for any charge distribution. The third term in Eq.~(\ref{epsExpansion}) is an additional term that contributes only when the net charge is non-zero, $q_l\neq 0$. Then, its contribution is also proportional to the second moment of the charge distribution.
This contribution is usually negative and may be significant in the case of macro-ions. In a solution of ions and dipoles of finite sizes, the ions thus contribute to the decrement of the dielectric constant, and this decrease depends linearly on the ionic concentration in the dilute limit, in agreement with experimental data.

The derivation within the DH approach as presented above offers only a minor modification to $\varepsilon$. The main effect comes from the hydration shell, and can only be obtained by treating the non-linear PB equation and will be presented next.

\section{Dipolar Poisson-Boltzmann}

Using the generalized version of the (non-linear) PB equation enables us to take into account the individual dipoles (together with the ions), instead of the medium constant dielectric background of the ``primitive model"~\cite{primitive}. This approach is called the \emph{Dipolar Poisson-Boltzmann} (DPB)~\cite{DPB}. There are three types of charges in the DPB model: permanent dipoles that can be conveniently modeled as  pairs of opposite charges ($\pm e$) with a small intra-pair distance $\mathbf{b}$, positive ions ($e$) and negative ones ($-e$), where $e$ is the electron charge. Note that throughout the remaining of this paper we consider only monovalent ions, $q_l=\pm e$, but the model can easily be generalized to any multi-valency.
The charges are free to move in the solution, whereas an additional fixed charge distribution resides on the boundary, and does not appear explicitly in the equation for the bulk. The various charge and dipole distributions can be written as:
\begin{eqnarray}
\label{dipole_length}
 \rho_d(\Omega; \rbf) & \approx &  e\mathbf{b} \cdot \nabla\delta(\rbf) \equiv \mathbf{p}_0 \cdot \nabla\delta(\rbf)
\nonumber\\
 \rho_+(\rbf) & = & e\delta(\rbf)
\nonumber\\
 \rho_-(\rbf) & = & -e\delta(\rbf),
\end{eqnarray}
where ${\bf p}_0=e{\bf b}$ is the individual dipole moment of each permanent dipole. Inserting $\rho_d(\Omega; \rbf), \rho_+(\rbf), \rho_-(\rbf)$ into Eq.~(\ref{expfullpb}) yields:
\begin{eqnarray}
\label{expfullpb_dpb}
 -\varepsilon_0\nabla^2\psi(\rbf) &  = &
   \Lambda_d \left<\rho_d(\Omega; -\rbf) \otimes \exp\left[ -\beta \rho_d(\Omega; \rbf)  \otimes \psi(\rbf)\right]\right>
\nonumber\\
& + &  \Lambda_s \left<\rho_+(-\rbf) \otimes \exp\left[-\beta  \rho_+( \rbf) \otimes \psi(\rbf\right]\right>
\nonumber\\
& + & \Lambda_s \left<\rho_-( -\rbf) \otimes \exp\left[ -\beta \rho_-( \rbf) \otimes \psi(\rbf)\right] \right>.\nonumber\\
\end{eqnarray}
There are three integrals to evaluate. The last two are very simple, because they are a convolution with a Dirac $\delta$-function: $\rho_\pm(\Omega; \rbf) \otimes \psi(\rbf) = \pm e \psi(\rbf)$, and $\rho_\pm(\Omega;-\rbf) \otimes \exp[ \mp\beta e \psi(\rbf)] = \pm e\exp[ \mp \beta e \psi(\rbf)]$. These terms give us a charge contribution just as in the standard PB equation. The more interesting part comes from the first term, for which the orientation averaging is non-trivial. The spatial integral can be solved by integration by parts. The result, before integrating over all possible orientations of the dipole $\mathbf{p}_0$ is $\langle \Lambda_d \mathbf{p}_0 \cdot \nabla \left[\exp \left(\beta \mathbf{p}_0 \cdot \nabla \psi(\rbf) \right)\right] \rangle_\Omega $. In order to calculate the integral over the spatial angles we have the freedom to choose any coordinate system we wish. The easiest choice would be one where the electric field ${\bf E} = -\nabla \psi$  is aligned with the $\hat{z}$ axis:
\begin{eqnarray}
\label{dpb_dipole_dist1}
 & & \langle \Lambda_d \mathbf{p}_0 \cdot \nabla {\rm e}^{- \beta \mathbf{p}_0 \cdot {\bf E} } \rangle_\Omega   =
\nonumber\\
& & \frac{1}{4\pi}\int_{-1}^{1} {\rm d}(\cos\theta)\, \int_{0}^{2\pi} {\rm d}\varphi\, \Lambda_d \mathbf{p}_0 \cdot \nabla {\rm e}^{-\beta p_0 E \cos\theta},
\end{eqnarray}
where $E = |{\bf E}|$ and $p_0=|{\bf p}_0|$. The vector $\mathbf{p}_0=(p_{0x},p_{0y},p_{0z})$ in spherical coordinates is equal to $\mathbf{p}_0 = p_0( \sin\theta \sin\varphi, \sin\theta \cos\varphi ,\cos\theta)$. From symmetry it is evident that the ${p}_{0x}$ and ${p}_{0y}$ contributions (in the $\hat{x}$ and $\hat{y}$ directions, respectively) equals to zero as we integrate over the solid angle $\Omega=(\theta,\phi)$. The only non-zero contribution comes from ${p}_{0z}$ component (in the $\hat{z}$ direction), as can be understood in the following way. Since $\hat{z}$ is the direction of the electric field, and the dipole moment has no other preferred direction. The $\hat{z}$ component of the dipole moment $p_{0z}= p_0 \cos \theta$ is multiplied by the $\hat{z}$ component of the $\bf{E}$ field. Since we have chosen the $\hat{z}$ axis to be in the direction of the electric field, we can write the unit vector $\hat{z}$ as $\hat{z}={\bf E}/E=\ehat$.
Integrating Eq.~(\ref{dpb_dipole_dist1}) over the angle $\varphi$ yields:
\begin{eqnarray}
& & \big< \Lambda_d \mathbf{p}_0 \cdot \nabla {\rm e}^{-\beta \mathbf{p}_0 \cdot \bf{E}} \big>_{\Omega}
\nonumber\\
&= & \frac{1}{2}\Lambda_d p_0 \nabla \cdot \left[\ehat \int_{-1}^{1} {\rm d}\cos\theta \, \cos\theta \, {\rm e}^{ -\beta p_0 E \cos\theta}\right].
\end{eqnarray}
Defining the function ${\cal G}(u)$
\begin{equation}
 {\cal G}(u) =  \frac{1}{2}\int_{-1}^{1}{\rm d}x\,x{\rm e}^{ux}=\frac{ \cosh u}{u} - \frac{\sinh u}{u^2},
\end{equation}
we can write the DPB equation as~\cite{DPB}:
\begin{eqnarray}
\label{DPBeqn}
-\varepsilon_0 \nabla^2 \psi & =& n_d p_0 \nabla \cdot \left[\frac{\nabla \psi}{|\nabla \psi|} {\cal G}(\beta p_0 |\nabla \psi|)\right]
\nonumber\\
& -& 2 n_s e\sinh\left[ \beta e \psi(\rbf)\right],
\end{eqnarray}
where the fugacities $\Lambda_d$ and $\Lambda_s$ are replaced, respectively, by their mean-field values (the bulk densities), $n_d$ and $n_s$. Note that the function ${\cal G}(u)$ is related to the Langevin function $L(u)= \coth(u) - 1/u$ by ${\cal G}(u)=L(u)\sinh(u) /u$.

\subsection{Field around a point-like ion}

The DPB equation, Eq.~(\ref{DPBeqn}), is a mean-field equation, where the contributions of the dipoles and charged particles appear on two decoupled terms in the RHS of Eq.~(\ref{DPBeqn}). Therefore, the dielectric decrement that is seen in experiments  cannot be  explained directly from the DPB model. However, since the discrete nature of the medium is considered, the model allows for a non-uniform dielectric response.

We can see how the ions affect the dielectric constant by choosing a model where the ions are held at fixed positions in a dielectric medium. The dipoles can move around, and will be treated using the DPB equation with boundary conditions set by the ions. To simplify the model we assume that the distance between any two ions is very large (dilute salt limit), and calculate the dielectric constant around a single ion, while neglecting all other ions, i.e., $n_s=0$, in Eq.~(\ref{DPBeqn}).

Adding a source term $\rho_f(\rbf)$ to Eq.~(\ref{DPBeqn}) yields
\begin{eqnarray}
\label{firstDPB}
-\varepsilon_0 \nabla^2 \psi & = & n_d p_0 \nabla \cdot \left[\frac{\nabla \psi}{|\nabla \psi|} {\cal G}(\beta p_0 |\nabla \psi|)\right]  +   \rho_f(\rbf)\, , \nonumber\\
\end{eqnarray}
where the source term is to be taken later as a charge density of a point particle  at the origin, $\rho_f(\rbf) = e\delta(\rbf)$.
In terms of the electric field ${\bf E} = -\nabla \psi$, the above equation becomes:
\begin{eqnarray}
\label{dpbEaxct}
\varepsilon_0 \nabla \cdot {\bf E} = - n_d p_0 \nabla \cdot \left[\ehat {\cal G}(\beta p_0 E)\right] + e\delta(\rbf) \, .
\end{eqnarray}
An analytical solution of the above non-linear PDE is probably too difficult to obtain. The linear DH regime results in an effective dielectric constant: $\varepsilon_0+\varepsilon_1$ where
\begin{equation}
\label{epsilon1}
\varepsilon_1= \frac{n_d \beta p_0^2}{3} \, .
\end{equation}
The effective $\varepsilon_0+\varepsilon_1$  plays the same role as the vacuum permittivity, $\varepsilon_0$, and thus will not give any new insight. In order to find an analytical result that captures the interactions between dipoles and ions, the next order in the Taylor expansion of ${\cal G}$ should be taken into account:
\begin{eqnarray}
\label{ThirdOrdEeqn}
\varepsilon_0  \nabla \cdot {\bf E} \approx -\varepsilon_1 \nabla \cdot {\bf E} -\frac{ n_d \beta^3 p_0^4 }{30}\nabla \cdot [{\bf E} E^2] + e\delta(\rbf).
\end{eqnarray}
Denoting ${\bf E}_1$ as the solution of the linearized form of the above equation
\begin{eqnarray}
\label{dpb_dh}
(\varepsilon_0 + \varepsilon_1) \nabla \cdot {\bf E}_1=e\delta(\rbf),
\end{eqnarray}
${\bf E}_1$  is the known Coulomb field for a charged particle at the origin. In order to see the response of the system to an external electric field $\mathbf{E}_0$, we can choose any external boundary condition that would create such a field (for example, two large capacitor plates with fixed and opposite charges). The induced displacement field (see Fig.~\ref{fig1}), $\mathbf{D}_1=(\varepsilon_0+\varepsilon_1){\bf E}_1$, for this system is equal to:
\begin{eqnarray}
\label{D0Solution}
\mathbf{D}_1 = (\varepsilon_0 + \varepsilon_1) \mathbf{E}_0 + \frac{e}{4 \pi r^2}\mathbf{\hat{r}},
\end{eqnarray}
where the second term is simply the electrostatic field originating from a charge particle (Coulomb law).
Inserting Eqs.~(\ref{dpb_dh}) and (\ref{D0Solution}) into Eq.~(\ref{ThirdOrdEeqn}) and using the displacement field ${\bf D}=(\varepsilon_0 + \varepsilon_1){\bf E}$, yields:
\begin{eqnarray}
\label{Deqn4}
 \nabla \cdot \mathbf{D}- \nabla \cdot \mathbf{D}_1 = - \frac{\varepsilon_1\beta^2 p_0^2}{10 (\varepsilon_1 + \varepsilon_0)^3} \nabla \cdot [\mathbf{D} D^2 ].
\end{eqnarray}
Integrating the above equation and assuming that $\mathbf{D}$ is in the direction of $\mathbf{D}_1$, leads to the following equation for $D$:
\begin{eqnarray}
\label{Deqn}
D^3 + (D^*)^2[D - D_1]=0,
\end{eqnarray}
where $D^*$ is a crossover field defined as:
\begin{equation}
\label{D^*def}
D^* =  \frac{1}{\beta p_0}\sqrt{\frac{10(\varepsilon_0 + \varepsilon_1)^3}{\varepsilon_1}}.
\end{equation}
Equation~(\ref{Deqn}) is a 3$rd$ order equation in $D$ and can be solved analytically. It has only one real root:
\begin{eqnarray}
\label{DD^*}
\frac{D}{D^*}  & = & \left[\frac{D_1}{2D^*} + \sqrt{\frac{1}{27} + \left(\frac{D_1}{2D^*}\right)^2}\,\right]^{1/3}
\nonumber\\
& - & \left[-\frac{D_1}{2D^*} + \sqrt{\frac{1}{27} + \left(\frac{D_1}{2D^*}\right)^2}\,\right]^{1/3} \, ,
\end{eqnarray}
which can be is written in a scaling form $D = D^*h(D_1/D^*)$. It is also worthwhile noticing that though we focus here on specific boundary conditions of a point-like ion at the origin, the same approximate solution of the DPB, Eq.~(\ref{DD^*}), can be obtained for any boundary conditions. Thus, any analytical solution of the PB equation (e.g., Refs.~\cite{ExactPBsol1, ExactPBsol2, ExactPBsol3, ExactPBsol4}), can be recast by Eq.~(\ref{DD^*}) to give an approximate solution to the DPB problem.

\begin{figure} [ht!]
  \centering
    \includegraphics[width=0.45\textwidth]{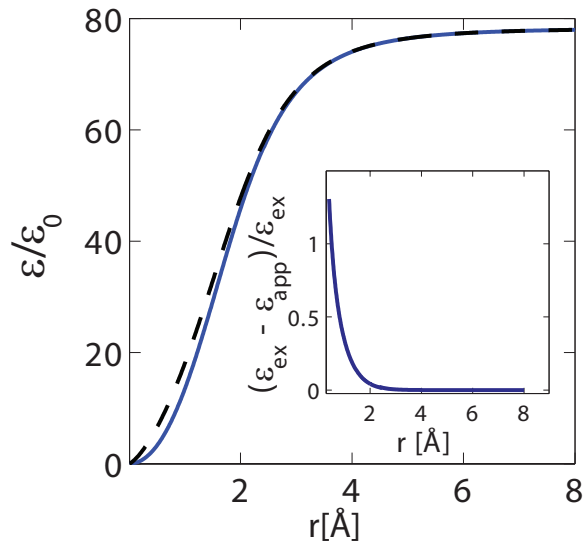}
\caption{\captionset{(color online) Approximated analytical solution (dashed line), Eq.~(\ref{eps_r_def}), and exact numerical solution (solid line), Eq.~(\ref{ExactEpsDpb}), for the dielectric constant of dipoles as function of the distance $r$ from a fixed point charge. In the inset the relative error between the two solutions is presented. The parameter values used are: $p_0 = 4.6$\,D, $T = 300$\,K and $n_s = 1$\,M. }}
\label{fig2}
\end{figure}

If we differentiate both sides of Eq.~(\ref{Deqn}) with respect to $E$ we get:
\begin{eqnarray}
\label{eps_r_def}
\varepsilon(r) & = & \frac{\varepsilon_0 + \varepsilon_1}{3 h^2\left(D_1(r)/D^*\right) + 1},
\end{eqnarray}
where $\varepsilon = \partial D / \partial E\Big|_{E_0=0}$ and $h(D_1/D^*)$ is obtained from Eq.~(\ref{DD^*}). The ratio between $D_1(\rbf)$ and $D^*$ can be expressed as:
\begin{eqnarray}
\label{D0D^*ratio}
\frac{D_1}{D^*} = \sqrt{\frac{\varepsilon_1}{10(\varepsilon_0+\varepsilon_1)}}\left(\frac{l_h}{r}\right)^2,
\end{eqnarray}
where we define a new length,
\begin{equation}
\label{lh}
l_h = \sqrt{\BejLen b} \, ,
\end{equation}
that characterizes the spatial behavior of the dielectric field, $\varepsilon(r)$, in terms of the Bjerrum length, $\BejLen = \beta e^2/4\pi(\varepsilon_0+\varepsilon_1)$ and the dipolar length, $b$, Eq.~(\ref{dipole_length}). The length $l_h$ can be thought of as the thickness of the hydration layer within our model since it describes a shell of dipoles surrounding an ion that are affected by it. Far away from the shell ($r \gg l_h$), we expect the dielectric constant to be equal to the bulk dielectric constant, $\varepsilon_0 + \varepsilon_1$. The leading term in $h(D_1/D^*)$ for large distances is $h(D_1/D^*) \approx  D_1/D^*$, thus the dielectric constant equals:
\begin{eqnarray}
\varepsilon (r \gg l_h) \approx \varepsilon_0 + \varepsilon_1 - \frac{1}{30} \varepsilon_1 \left(\frac{l_h}{r}\right)^4.
\end{eqnarray}
In the charge vicinity, $r\ll l_h$, the leading term of $h$ is $h(D_1/D^*)\approx (D_1/D^*)^{1/3}$, which yields:
\begin{eqnarray}
\label{small_r_eps}
\varepsilon (r \ll l_h) \approx \frac{10^{1/3}}{3}(\varepsilon_0 + \varepsilon_1) \left(\frac{\varepsilon_0 + \varepsilon_1}{ \varepsilon_1}\right)^{1/3} \left(\frac{r}{l_h}\right)^{4/3}.\nonumber\\
\end{eqnarray}
The approximate analytical behavior of $\varepsilon(r)$ around a point-like particle is illustrated on Fig.~\ref{fig2} (dashed line). Very close to the charged particle the external electric field does not affect the dipoles, leading to zero contribution to the dielectric constant. As the distance $r$ grows, $\varepsilon(r) \sim r^{4/3}$ within the hydration layer. Farther away from the charge, $r \gg l_h$, $\varepsilon(r)$ asymptotes the bulk value of the dielectric constant.

\begin{figure} [ht!]
  \centering
    \includegraphics[width=0.45\textwidth]{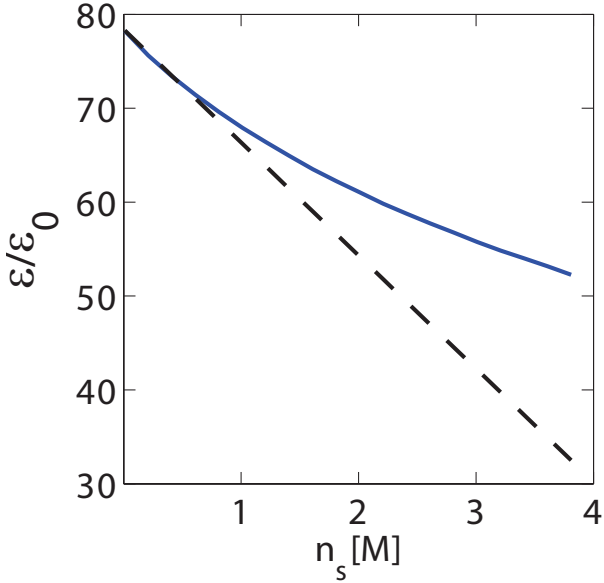}
\caption{\captionset{(color online) Numerical evaluation (solid line) of the average dielectric constant, according to Eq.~(\ref{avg_eps_int}), and approximate solution (dashed line) for the dilute limit, Eq.~(\ref{epsDPBtotal}), for ionic concentration of up to $4$\,M. The values of the parameters are $\varepsilon_1=77\varepsilon_0$ and $l_h = 2.7$\AA.}}
\label{fig3}
\end{figure}

We may calculate the effective dielectric constant and extract the average decrement.
The correction term is given by:
\begin{eqnarray}
\label{third_order_sol}
\Delta\varepsilon(r) & = & \varepsilon - \varepsilon_0 - \varepsilon_1 = -\frac{3(\varepsilon_0+\varepsilon_1)}{3 + 1/h^2\left(D_1(r)/D^*\right)}.
\end{eqnarray}
In a dilute solution the ions are unaffected by each other, and the effective dielectric constant can be evaluated by averaging $\varepsilon$ in a sphere around each ion. The diameter of this sphere is set by equating it to the distance between nearest-neighbors residing on an equivalent cubic lattice. For 1:1 salt with ionic density of $n_s$, the radius of the sphere equals to  $R=(2n_s)^{-1/3}/2$, and

\begin{eqnarray}
\label{avg_eps_int1}
\varepsilon(n_s)=\varepsilon_0+\varepsilon_1+\langle\Delta{\varepsilon}\rangle \, ,
\end{eqnarray}
where
\begin{eqnarray}
\label{avg_eps_int}
\langle\Delta{\varepsilon}\rangle =  \frac{3}{4\pi R^3} \int {\rm d}^3 \rbf \, \Delta \varepsilon(\rbf).
\end{eqnarray}
We can evaluate the integral in Eq.~(\ref{avg_eps_int}) numerically for different values of $n_s$ and calculate the dielectric constant $\varepsilon(n_s)$.
The result of the numerical integration is plotted in Fig.~\ref{fig3} for ionic concentrations of up to $4$\,M, and compared with the linear decrement approximation of Eq.~(\ref{epsDPBtotal}) that is presented next. For concentration above $1$\,M a substantial deviation from linearity can be seen.

First we note that for pure water at room temperature, $T=300\,$K, and for dipolar moment $p_0 =1.8$\,D and density $n_d=55$\,M, the obtained value of $\varepsilon_1$ is $11.1\varepsilon_0$. Hence, $\varepsilon_w=\varepsilon_0+\varepsilon_1 \simeq 12.1\varepsilon_0$.
Note that this value is much smaller than the measured one, $\varepsilon_w=78\varepsilon_0$. This is not surprising since the model uses a dilute gas approximation, which does not capture
the correlation effects of dense liquid water. To overcome this problem, the dipole moment $p_0$ is treated as a fitting parameter, and is set to be $p_0=4.6$\,D, in order to match the value of pure water, $\varepsilon_w=78\varepsilon_0$. We also note that for water with $\BejLen = 7 \AArm$ and $b = 1 \AArm$, the size of the hydration shell equals $l_h \simeq 2.6 \AArm$, which is comparable to the size of water molecules.

In the very dilute ionic limit, $R \rightarrow \infty$, the integration in Eq.~(\ref{avg_eps_int}) can be evaluated analytically. We expressed it in term of the $\gamma$ defined in Eq.~(\ref{DecrementForm}):
\begin{eqnarray}
\label{epsDPBtotal}
\varepsilon&=&\varepsilon_0 + \varepsilon_1+\gamma n_s \, ,\nonumber\\
\gamma &=& -\eta (\varepsilon_0 + \varepsilon_1)  \left(\frac{\varepsilon_1}{\varepsilon_0 + \varepsilon_1}\right)^{3/4} l_h^3 \, ,
\end{eqnarray}
where $\eta$ is a dimensionless numerical pre-factor,
\begin{eqnarray}
\eta  =  24 \cdot 10^{-3/4}  \int_0^\infty \frac{3x^{-5/2}}{3+1/h^{2}(x)} \,{\rm d} x \approx  13.87.
\end{eqnarray}

\subsection{Numerical solution of the DPB equation}
The results in the last section were based on the assumption that expanding ${\cal G}$ to third order will be sufficient to capture the interaction between the ion in the origin and the dipoles. However, in the vicinity of the ion, the electric field diverges and, thus, the approximation may not be valid anymore.

Let us extend our results and examine a numerical solution of the full DPB equation in comparison with the approximate solution of Eq.~(\ref{third_order_sol}).
The DPB equation, Eq.~(\ref{DPBeqn}), can be written in terms of the displacement field ${\bf D}$:
\begin{eqnarray}
\label{fullDPBsol1}
\nabla \cdot \left\{\frac{\varepsilon_0}{\varepsilon_0 + \varepsilon_1} {\bf D} + n_d p_0 \left[\ehat {\cal G}\left(\frac{\beta p_0}{\varepsilon_0 + \varepsilon_1} D\right)\right]\right\} = \nabla \cdot \mathbf{D}_1. \nonumber\\
\end{eqnarray}
By the same argument used in Sec.~III.A, we integrate both sides of Eq.~(\ref{fullDPBsol1}) and get a nonlinear equation:
\begin{eqnarray}
\label{fullDPBsol2}
\frac{\varepsilon_0}{\varepsilon_0 + \varepsilon_1} D + n_d p_0 {\cal G}\left(\frac{\beta p_0}{\varepsilon_0 + \varepsilon_1} D\right)- D_1=0.
\end{eqnarray}
There are many ways to solve numerically such nonlinear equations, and we chose the fast converging Newton-Raphson method~\cite{NumericalRecepies}. The approximate analytical solution of Eq.~(\ref{DD^*}) was chosen as the starting point for the numerical iterative process. The dielectric constant is derived by differentiating Eq.~(\ref{fullDPBsol1}) with respect to $E_0$, and then substituting the numerical solution for $D(r)$:
\begin{eqnarray}
\label{ExactEpsDpb}
\varepsilon(r) & = & \frac{(\varepsilon_0 + \varepsilon_1)^2}{\varepsilon_0 + 3\varepsilon_1 {\cal G}'\left(\frac{\beta p_0}{\varepsilon_0 + \varepsilon_1} D(r)\right)},
\end{eqnarray}
where ${\cal G}'(u) = {\rm d} {\cal G}(u)/{\rm d} u$ is equal to:
\begin{eqnarray}
{\cal G}'(u) = \frac{\sinh u}{u}\left(1+\frac{2}{u^2}\right) - \frac{2\cosh u}{u^2}.
\end{eqnarray}

In Fig.~\ref{fig2} we compare the exact (numerical) and the approximate results for the dielectric constant. As expected, in the vicinity of the ion the electric field is strong and the approximation deviates from the numerical solution (see inset of Fig.~\ref{fig2}), though both calculations show that the dielectric constant goes to zero at the origin. For distances $\sim 2 \, {\rm \AA}$, there is less that $ 5\% $ difference between the approximate and exact (numerical) solutions; namely, our approximate solution works rather well.

After showing the validity of the approximate solution (Fig.~\ref{fig2}) we can extend the DPB formalism to incorporate other physical details. In particular, three cases are examined: finite size ions, binary mixtures of dipolar solvents and polarizability effects.

\subsection{Field around a finite-size ion}

\begin{figure} [ht!]
  \centering
    \includegraphics[width=0.48\textwidth]{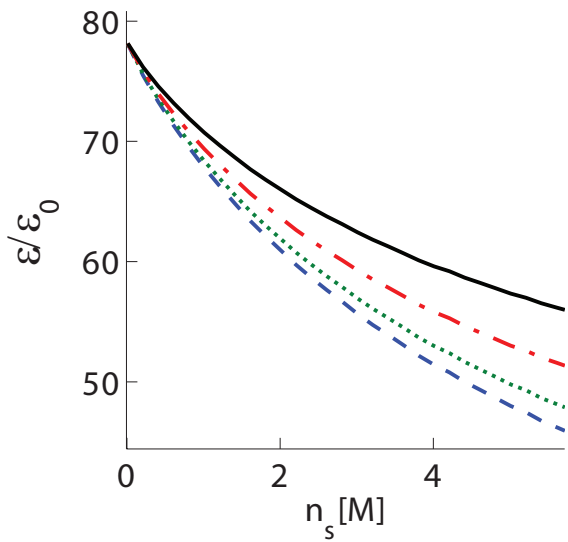}
\caption{\captionset{(color online) The spatial average dielectric constant $\varepsilon/\varepsilon_0 = \langle \varepsilon \rangle/\varepsilon_0$, Eq.~(\ref{finite_size_eps_integral}), as a function of salt concentration for four ionic radii: $2$\,\AA\,(black, solid line), $a=1.5$\,\AA \,(red, dot-dashed line), $a=1$\,\AA \,(green, dotted line) and $a=0.1$\,\AA\, (blue, dashed  line). The decrement is more pronounced for small ionic radii, where the ion size is much smaller than the hydration length, $l_h=2.7$~\AA}.
\label{fig4}}
\end{figure}

For ions with finite size, the solution of the DPB equation depends only on the local electrostatic field. Thus, the solution for a sphere-like particle is the same  as that of a point-like ion. If we neglect the inner dielectric properties of the ions, the only difference is in the calculation of the average dielectric constant, $\langle \varepsilon \rangle$. In case of finite-size ions, the averaging over the dielectric constant starts from the radius of the sphere, denoted by $a$. The upper limit, as in Eq.~(\ref{avg_eps_int}), is defined by the ionic concentration, $R = (2n_s)^{-1/3}/2$:
\begin{eqnarray}
\langle\varepsilon(n_s,a)\rangle = \frac{\int_a^R {\rm d}^{3} \rbf\, \varepsilon(\rbf)}{\frac{4\pi}{3}\left(R^3-a^3\right)}
\end{eqnarray}
Assuming that the ionic size $a$ is small compare to the hydration length $l_h$, the approximation of $\varepsilon(\rbf)$ for small distances, Eq.~(\ref{small_r_eps}), can be used to obtain a closed-form formula for $\langle\varepsilon(n_s,a)\rangle$, as a function of the expression obtained in Eq.~(\ref{avg_eps_int}) for zero size, $\langle \varepsilon(n_s,a=0)\rangle$:
\begin{eqnarray}
\label{finite_size_eps_integral}
\langle\varepsilon(n_s,a)\rangle & = & \frac{3}{4\pi\left(R^3-a^3\right)}\left(\int_0^R {\rm d}^{3} \rbf\, \varepsilon(\rbf)- \int_0^a {\rm d}^{3} \rbf\, \varepsilon(\rbf)\right)
\nonumber\\
& = & \frac{R^3}{R^3 - a^3}\langle\varepsilon(n_s,0)\rangle - \frac{3\int_0^a {\rm d}^{3} \rbf \,\varepsilon(\rbf)}{4\pi\left(R^3-a^3\right)}.
\end{eqnarray}
The assumption that $a \ll l_h$ implies also that $a \ll R$, so within this approximation only the first-order term in $a/R$ is taken into account. Using the relation between $R$ and $n_s$ yields the following approximation for the dielectric constant of finite-size ionic solution:
\begin{eqnarray}
\label{finite_size_eps}
& & \langle\varepsilon(n_s,a)\rangle  \approx  \langle\varepsilon(n_s,0)\rangle+ 16  \langle\varepsilon(n_s,0)\rangle a^3 n_s
\nonumber\\
& - & \frac{48}{13}  (\varepsilon_0 + \varepsilon_1) \left(\frac{10(\varepsilon_0 + \varepsilon_1)}{ \varepsilon_1}\right)^{1/3}l_h^{-4/3} a^{13/3} n_s.
\end{eqnarray}
As was seen for standard parameter values at room temperature, the hydration length equals to $l_h \simeq 2.6$\AA, and is quite comparable with size of large ions. Thus, Eq.~(\ref{finite_size_eps}) is valid only for very small ions~\cite{IonicRadii}. For larger $a$ we have to evaluate the full integral in Eq.~(\ref{finite_size_eps_integral}), using $\varepsilon(\rbf)$  from Eq.~(\ref{eps_r_def}). The results of a numerical integration are plotted in Fig.~\ref{fig4}. As expected, large ions cause a smaller decrement of the dielectric constant.

\subsection{Mixture of dipoles}
The solvent in the usual DPB theory, as well as in other PB generalizations, is water~\cite{DPB, DLVO, ExcludedVolume, PPB}. However, it can be interesting to investigate the behavior of other solvents as well as binary  mixtures of solvents~\cite{BinaryMixLowTemp, Mixture1, Mixture2, Mixture3, Mixture4, Mixture5}.

On the mean-field level, the dielectric constant of a mixture of solvents equals to the weighted average of the dielectric constants of each of the solvents, weighted by their relative volume fraction, as is appropriate from Eq.~(\ref{epsExpansion}).
Let us consider in more detail the DPB of a binary mixture of solvents, and derive its ``hydration length". This is the length scale that determines the dielectric decrement, as was shown in Eq.~(\ref{eps_r_def}).

The DPB equation for an A/B solvent mixture is a generalization of Eq.~(\ref{DPBeqn}) and reads:
\begin{eqnarray}
\label{DPBeqnMix}
\varepsilon_0 \nabla^2 \psi &=& -\rho_f(\rbf) + 2n _s e\sinh\left( \beta  e\psi\right)\nonumber\\
 &-& \phi n_d p_A \nabla \cdot \left[\frac{\nabla \psi}
{|\nabla \psi|} {\cal G}(\beta p_A |\nabla \psi|)\right]
\nonumber\\
&  - &(1-\phi) n_d p_B \nabla \cdot \left[\frac{\nabla \psi}
{|\nabla \psi|} {\cal G}(\beta p_B |\nabla \psi|)\right],
\end{eqnarray}
where $\phi_A=\phi$ is the volume fraction of the A solvent, $\phi_B=1-\phi$ is the volume fraction of the B solvent, and $p_A$ and $p_B$ are the dipole moments of the two solvents. We need to expand Eq.~(\ref{DPBeqnMix}) at least to 3${rd}$ order, because the 1$st$ order will simply give an effective average contribution to the dielectric constant. Removing the ionic part, $n_s=0$, and setting $\rho_f=e\delta(r)$ yield an equation with the same structural form of the DPB as in Eq.~(\ref{ThirdOrdEeqn}):
\begin{eqnarray}
\label{mix_dpb}
(\varepsilon_0+\frac{1}{3}\beta n_d \langle p^2 \rangle_\phi)\nabla \cdot {\bf E} = -\frac{1}{30} n_d \beta^3  \langle p^4 \rangle_\phi \nabla \cdot [{\bf E} E^2] + e\delta({\mathbf r}),
\nonumber\\
\end{eqnarray}
where $\langle\ldots\rangle_\phi$ denotes averaging by volume fraction, and the 2${nd}$ and 4${th}$ moments are:
\begin{eqnarray}
\langle p^2 \rangle_\phi & = & \phi p_A^2 + (1-\phi) p_B^2,
\nonumber\\
\langle p^4 \rangle_\phi & = & \phi p_A^4 + (1-\phi) p_B^4.
\end{eqnarray}
From the analogy with the DPB equation for a single solvent, Eq.~(\ref{D0D^*ratio}), we get the following hydration length:
\begin{eqnarray}
\label{l_h_mix}
l_h = \sqrt{l_B b} \frac{\sqrt{\langle p^4 \rangle_\phi}}{\langle p^2\rangle_\phi},
\end{eqnarray}
where the effective (averaged) Bjerrum length $\BejLen$ is:
\begin{eqnarray}
\label{l_b_mix}
\BejLen & = & \frac{\beta e^2}{4\pi (\varepsilon_0 + \frac{1}{3} n_d \beta \langle p^2 \rangle_\phi)},
\end{eqnarray}
and similarly $b=\sqrt{\langle p^2 \rangle_\phi/e^2}$.

\begin{figure} [ht!]
  \centering
    \includegraphics[width=0.45\textwidth]{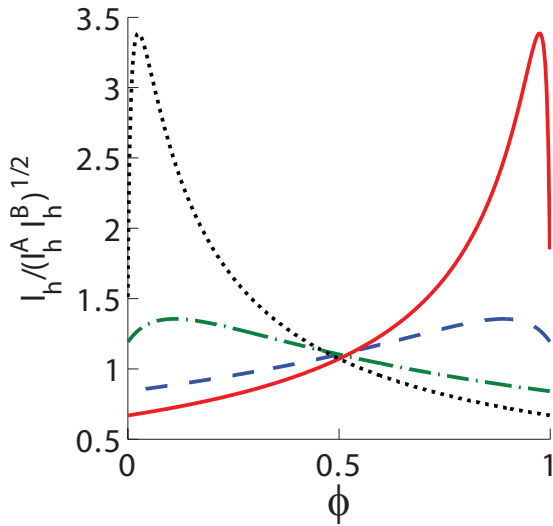}
\caption{\captionset{\label{fig5}(color online) The ratio between the hydration length of a binary mixture of dipoles, $l_h$,
and the geometric mean of the pure A and B $l_h$: $\sqrt{l_h^A l_h^B}$, as a function of the relative A/B concentration, $\phi$. Because of the normalization factor, $l_h(\phi=0)/\sqrt{l_h^A l_h^B}=(p_B/p_A)^{1/4}$ and $l_h(\phi=1)/\sqrt{l_h^A l_h^B}=(p_A/p_B)^{1/4}$
Four different mixtures are considered with relative dipole moments: $p_B = 0.2p_A$ (red solid line), $p_B=0.5p_A$ (blue dashed line),
$p_B=2p_A$(green dash-dot line) and $p_B=5p_A$ (black dotted line).
}}
\end{figure}

In Fig.~\ref{fig5} the effective hydration length, $l_h$, is plotted as a function of the A/B volume fraction $\phi$, for different ratios of $p_B/p_A$. In the limits $\phi \to 0$ and $\phi \to 1$, we get the single-solvent hydration length, as expected.
The trend however is not linear, and the larger of the two dipole moments becomes the dominant one rapidly as its concentration increases. Even a small volume fraction of highly polar molecules can change the hydration length greatly. The dielectric decrement at the dilute limit is proportional to the hydration length, Eq.~(\ref{epsDPBtotal}), and can be manipulated by changing the relative A/B volume fraction.

\subsection{Polarizability effects: A spring-dipole model}
To conclude this section, we consider a variation of the DPB model that incorporates polarizability in addition to permanent dipole moment~\cite{PPB}. For polarizable molecules, the external electric field induces a dipole moment and changes the internal charge distribution. In the general description of the charge distribution used so far, Eq.~(\ref{chargdist}), we allowed only for rotations and translations of the same charge distribution, but for polarizable media, an additional degree of freedom exists. Another variant model was introduced recently in Ref.~\cite{Podgornik2012}, where the model included polarizable counterions instead of polarizability of the dipolar molecules as is done here.

For simplicity, we limit our discussion only to spring-like dipoles, where two opposite charges are connected with a variable length spring, while the ions are taken as point-like. Taking the dipole length, $b$, as the new degree of freedom, the free energy can be written as a sum over the electric and elastic free energies of $N_d$ dipoles. The elastic contribution due to spring deformation is equal to:
\begin{eqnarray}
F_{\rm elastic}= \frac{ \kappa}{2} \sum_{l=1}^{N_d}(b_l-b_0)^2 \, ,
\end{eqnarray}
where $\kappa$ is the spring constant, $b_0$ is the rest length, and $b_l$ is the length of the $l^{th}$ spring-dipole. The dipole moment $p_0 = e b_0$ plays the role of the permanent dipole moment, because it exists even in the absence of an external field. For reasons that will become apparent shortly, the dipole moment that corresponds to the polarizability equals to $p_\alpha = \sqrt{ e^2/\beta \kappa}$. Thus, the elastic energy can be recast as:
\begin{eqnarray}
\label{spring_energy}
\beta F_{\rm elastic}= \sum_{l=1}^{N_d}\frac{(p_l-p_0)^2}{2 p_\alpha^2} \, .
\end{eqnarray}
Once adding the elastic term, Eq.~(\ref{spring_energy}), to the grand-canonical partition function, Eq.~(\ref{GC}), the Euler-Lagrange equation becomes:
\begin{eqnarray}
\label{DPPB}
-\varepsilon_0 \nabla^2 \psi & = &  n_d  \nabla \cdot \left[\frac{\nabla \psi}{|\nabla \psi|} \langle p {\cal G}(\beta p |\nabla \psi|)\rangle_p\right]
\nonumber\\
& - & 2 n_s e\sinh\left[ \beta e \psi(\rbf)\right],
\end{eqnarray}
where $\langle\ldots\rangle_p$ denotes averaging over the dipole moment $p$:
\begin{eqnarray}
\langle f(p) \rangle_p = \frac{ \int_0^{\infty} {\rm d}p\,f(p){\rm e}^{-(p-p_0)^2/2p_\alpha^2}}{\int_0^{\infty} {\rm d}p\,{\rm e}^{-(p-p_0)^2/2p_\alpha^2}}.
\end{eqnarray}
Equation~(\ref{DPPB}) has the same structure as the standard DPB equation (\ref{DPBeqn}), where the function $\cal{G}$ is
replaced with a more complicated function that has no simple analytical form. Nevertheless, it can be expanded to 3$rd$  order in a Taylor series:
\begin{eqnarray}
\label{dppb_third_ord}
& & (\varepsilon_0+\frac{1}{3}\beta n_d \langle p^2 \rangle_p)\nabla \cdot {\bf E}  =
\nonumber\\
& - & \frac{1}{30} n_d \beta^3  \langle p^4 \rangle_p \nabla \cdot [{\bf E} E^2] + e\delta(\rbf).
\end{eqnarray}
Note that Eq.~(\ref{dppb_third_ord}) is exactly the same as Eq.~(\ref{mix_dpb}), with a different interpretation of the averaging operation. The averages in Eq.~(\ref{dppb_third_ord}) can be expressed using the error function (${\rm erf}$):
\begin{eqnarray}
\langle p^2\rangle_p & = & (p_\alpha^2+p_0^2)
 + \sqrt{\frac{2}{\pi}}\frac{p_\alpha p_0 {\rm e}^{-p_0^2/2p_\alpha^2}}{1+{\rm erf}(p_0/\sqrt{2}p_\alpha)},
\end{eqnarray}
and
\begin{eqnarray}
\langle p^4 \rangle_p  &= & p_0^4 + 6p_0^2p_\alpha^2+3p_\alpha^4
 + \sqrt{\frac{2}{\pi}}\frac{(p_0^3 p_\alpha+5p_\alpha^3p_0) {\rm e}^{-p_0^2/2p_\alpha^2}}{1-{\rm erf}(p_0/\sqrt{2}p_\alpha)}.
\nonumber\\
\end{eqnarray}
Consequently, the dielectric constant in the mean-field level is equal to:
\begin{eqnarray}
\label{eps_dppb}
\varepsilon = \varepsilon_0 + \frac{1}{3} n_d \beta \left(p_\alpha^2+p_0^2 +
 \sqrt{\frac{2}{\pi}}\frac{p_\alpha p_0 {\rm e}^{-p_0^2/2p_\alpha^2}}{1+{\rm erf}(p_0/\sqrt{2}p_\alpha)}\right).
\nonumber\\
\end{eqnarray}

We can connect now $p_\alpha$ to the polarizability, $\alpha$, defined as the relation between the
induced dipole moment and the external electric field $E_0$, $p=p_0+\alpha E_0$. In the spring-dipole model
we can extract this relationship via minimization of the free energy:
\begin{eqnarray}
\beta F & = & \beta pE_0 - \frac{(p-p_0)^2}{2 p_\alpha^2}  ,
\nonumber\\
\frac{\partial F}{\partial p} & = &  E_0 - \frac{(p-p_0)}{\beta p_\alpha^2}=0  ,
\nonumber\\
p & = & p_0+  \beta p_\alpha^2 E_0  ,
\nonumber\\
p_\alpha & = & \sqrt{\frac{\alpha}{\beta}}  .
\end{eqnarray}

\begin{figure} [ht!]
  \centering
    \includegraphics[width=0.45\textwidth]{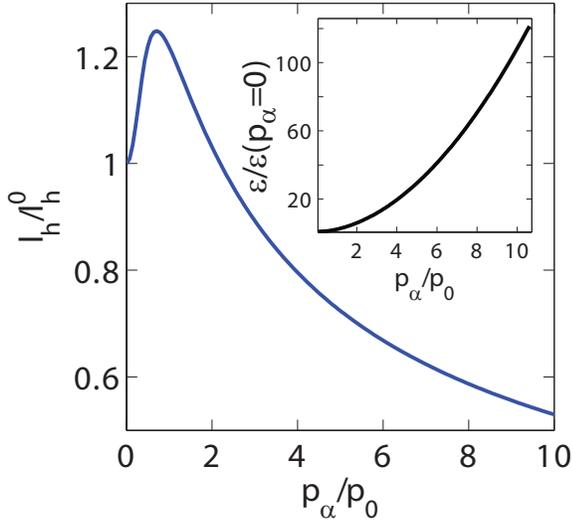}
\caption{\captionset{\label{fig6}(color online) The relative hydration length, $l_h/l_h^0$,  as a function of the relative polarizability, $p_\alpha/p_0$, where $l_h^0=l_h(p_\alpha{=}0)$ and $p_\alpha=0$ is the permanent-dipole only case. In the inset, the relative dielectric constant, $\varepsilon/\varepsilon(p_\alpha{=}0)$ is plotted as function of $p_\alpha$.}}
\end{figure}

Indeed, the spring-dipole model predicts a linear dependence of the induced dipole on the external field. However, taking into account the thermodynamical average leads to a more complex behavior. Both the dielectric constant, Eq.~(\ref{eps_dppb}), and the hydration length [according to Eq.~(\ref{l_h_mix})] are plotted in Fig.~\ref{fig6} as a function of the ratio $p_\alpha/p_0$. As expected, the dielectric constant increases with the polarizability, while the hydration length decreases. The treatment here is similar to the one done by Frydel~\cite{PPB}, but is cast in our general DPB framework.

\section{One-Loop Expansion of the DPB}
In the previous section we used the DPB equation on a mean-field level to calculate the decrement of the dielectric constant. In order to capture the interactions between  ions and the dielectric medium we treated the ions as fixed charges, and not as mobile particles in thermodynamical equilibrium. In this section we present a more complete model that goes beyond mean-field. The dielectric decrement is calculated in a complete statistical mechanical manner, by a direct derivation from the grand-canonical partition function. Since the partition function is a functional integral we approximate it using the \emph{loop expansion} method.

The method of loop expansion~\cite{LoopExp1,LoopExp2} is a special resummation of Feynman diagrams corresponding to a systematic saddle-point expansion. It is used in Quantum Field Theory (QFT) as a systematic way of calculating quantum-mechanical amplitudes of different physical processes. The amplitude is written as a functional integral where each field configuration is weighted by its classical action, and the diagrams provide an elegant way of expanding the solution as a function of a small parameter. The analogy between functional integrals of the partition function in statistical mechanics and path integrals of QM amplitudes~\cite{LoopExp2,FeynmanHibbs} enables us to use similar tools in our calculation.

We start with a general functional integral of the form, Eq.~(\ref{GC}):
\begin{eqnarray}
\label{GeneralPathInt}
\Xi = \int {\cal D} \phi(\rbf) {\rm e}^{- \beta F[\phi(\rbf)]},
\,\,\,\,\,\,\,\,
\end{eqnarray}
where $\phi(\rbf)$ is a field, $F$ is a functional of the field $\phi(\rbf)$. The first-order correction of Eq.~(\ref{GeneralPathInt}) is given by the one-loop order in the expansion~\cite{ByondPB}:
\begin{eqnarray}
\label{PathIntApprox}
\Xi &\simeq&  {\cal N}\exp \left\{ -  \beta F[\phi_{\rm MF}(\rbf)] - \frac{1}{2} \ln \left[ \det \left(\frac{\delta^2 F[\phi(\rbf)]}{\delta \phi(\mathbf{r'}) \delta\phi(\rbf)}\right)\right]\right\},
\nonumber\\
\end{eqnarray}
where $\phi_{\rm MF}$ is the solution of the mean-field DPB equation (as was presented in Sec. III) and  ${\cal N}$ is a normalization constant.
Since we are interested in the bulk value of the dielectric constant, the DPB solution is simply $\phi_{\rm MF}=0$.

For the DPB model, the functional $F$ and its second functional derivative (the Hessian), $F^{(2)}=\delta^2 F[\phi(\rbf)]/\delta \phi(\mathbf{r'}) \delta\phi(\rbf)$,  are given by:
\begin{eqnarray}
\label{DPBfunctinal}
& - \beta F = &\int {\rm d}^3\rbf\,\bigg\{  -\frac{\varepsilon_0\beta}{2}[\nabla\phi(\rbf)]^2
+2\Lambda_s \cos \left[\beta e \phi(\rbf)\right]
\nonumber\\
 & +&  \Lambda_d \int \frac{{\rm d}^2 \Omega}{4\pi}\,{\rm e}^{ i \beta {\bf p}_0 \cdot \nabla \phi(\rbf)}\bigg\} \, ,
\end{eqnarray}
and
\begin{eqnarray}
\label{DPBfunctinal2}
F^{(2)} & = &- \varepsilon_0 \nabla^2\delta( \mathbf{r-r'}) +  2\Lambda_s \beta e^2\cos \left[ \beta e \phi(\rbf)\right] \delta( \mathbf{r-r'})
\nonumber\\
 &+&  \Lambda_d \beta\int {\rm d}^3 \mathbf{r''} \int \frac{{\rm d}^2 \Omega}{4\pi}\,  {\rm e}^{i \beta {\bf p}_0 \cdot \nabla \phi(\mathbf{r''})}
\nonumber\\
&\times& \left[\mathbf{p}_0 \cdot \nabla  \delta( \mathbf{r-r''})\right]
\left[\mathbf{p}_0 \cdot \nabla  \delta( \mathbf{r'-r''})\right].
\end{eqnarray}

The determinant of any operator is equal to the product of its eigenvalues.  Evaluation of the logarithm of this determinate leads to divergences. Fortunately, we are not interested in the value of the grand-canonical partition function itself, but only in its derivatives at $\phi=0$. Keeping this in mind, we can use a general formula for matrices and operators that depend on a parameter $\alpha$:
\begin{eqnarray}
\label{TraceLogDer}
\frac{\partial \ln(\det A)}{\partial \alpha} = \int {\rm d}^3 \rbf  \int {\rm d}^3 \mathbf{r'} A^{-1}(\rbf,\mathbf{r'}) \frac{\partial A}{\partial \alpha}.
\end{eqnarray}
Using Eq.~(\ref{TraceLogDer}) allows us to avoid calculating the determinant explicitly. Instead, we need to know the inverse of the $F^{(2)}$  operator (the Green's function) at $\phi_{\rm MF}=0$. It is denoted by $g$ and given by:
\begin{equation}
g(\mathbf{r,r'}) = \frac{1}{4 \pi \beta (\varepsilon_0 + \varepsilon_1)} \frac{{\rm e}^{-\DebFreq |\mathbf{r-r'}|}}{ |\mathbf{r-r'}|},
\end{equation}
where $\varepsilon_1 = \beta p_0^2 n_d/3$ was defined in Eq.~(\ref{epsilon1}) and $\DebFreq$ is the inverse Debye length:
\begin{eqnarray}
\label{xiDef}
\DebFreq = \frac{1}{\DebLen} = \sqrt{\frac{2 n_s \beta e^2}{\varepsilon_0 + \varepsilon_1}}.
\end{eqnarray}

The dielectric constant can be derived as a thermodynamical average from the grand-canonical
partition function.
The dielectric response is obtained by taking the second functional derivative of the free
energy $F$ with respect to the electrostatic field $\mathbf{E}$.
The dielectric constant for an isotropic homogeneous medium is given by:
\begin{eqnarray}
\label{epsilon_general}
\varepsilon & = & \int {\rm d}^3 \rbf \frac{\delta^2 F}{\delta E_i(\rbf) \delta E_i(\rbf^\prime)}.
\end{eqnarray}
Due to isotropy, the direction of the electric field $E_i$ is arbitrary, and  translational invariance implies that the second functional derivative is only a function of $\rbf-\rbf^\prime$.
On a mean-field level, this results in $\varepsilon_1 = \beta n_d p_0^2/3$ that is a
function of the bulk concentration, $n_d$. Hence, both the dielectric constant and the densities
have to be calculated consistently up to 1$st$ order in the loop expansion.

The average number of particles can be derived from the grand-canonical partition function as:
\begin{eqnarray}
\langle N \rangle & = & \Lambda \frac{\partial \ln\Xi}{\partial \Lambda}.
\end{eqnarray}
This equation is valid both for the dipole number $N_d$, and for the charge number $N_s$, with corresponding $\Lambda_d$ and $\Lambda_s$. In the mean-field approximation $\Lambda_d = n_d$ and $\Lambda_s = n_s$. The one-loop correction is given by:
\begin{eqnarray}
\label{DeltaNs}
n_s & = &  \Lambda_s + \left.\frac{\Lambda_s}{4V}\frac{\partial \ln \left[\det (F{(2)})\right]}{\partial \Lambda_s}\right|_{\phi_{\rm MF}=0}
\nonumber\\  &  = & \Lambda_s + \frac{\Lambda_s}{2V} (\beta e)^2 \int {\rm d}^3 \rbf \, \int {\rm d}^3 \mathbf{r'} \,g(\mathbf{r,r'}) \delta( \mathbf{r-r'})
\nonumber\\
& = &\Lambda_s + \frac{\Lambda_s}{2}(\beta e)^2 g(0) .
\end{eqnarray}
The correction for $n_d$
is calculated in a similar manner, and results in:
\begin{eqnarray}
\label{DeltaNd}
n_d =    \Lambda_d -\frac{\beta \Lambda_d}{2} \frac{\beta p_0^2}{3} \nabla^2 g(0) \, .
\end{eqnarray}
The correction terms for the fugacities depend on the diverging Green's function value $g(r)$ at $r \rightarrow 0$. In order to avoid this divergence we need to consider a minimal cutoff distance $a$ between particles. Alternatively, one can use self-energy regulation techniques~\cite{SelfEnergy}. The cutoff distance $a$  corresponds to a maximal wavenumber $k_{\rm max} = 2 \pi/a$. By considering the solution in Fourier space, the value of the Green's function and its Laplacian at $r \rightarrow 0$ are approximated by:
\begin{eqnarray}
\label{Gvalue}
2\pi^2\beta (\varepsilon_0+\varepsilon_1) g(0) & = & k_{\rm max} - \DebFreq\tan^{-1}\frac{k_{\rm max}}{\DebFreq}
\nonumber\\
2\pi^2\beta (\varepsilon_0+\varepsilon_1)\nabla^2 g(0) & = &
 - \frac{k_{\rm max}^3}{3} +k_{\rm max}\DebFreq^2 - \DebFreq^3\tan^{-1}\frac{k_{\rm max}}{\DebFreq}.
\nonumber\\
\end{eqnarray}
Substituting Eq.~(\ref{Gvalue}) into Eq.~(\ref{DeltaNs}) and Eq.~(\ref{DeltaNd}) we can write the first-order correction to the fugacities $\Lambda_d$ and $\Lambda_s$:
\begin{eqnarray}
\label{fugacity_one_loop}
\Lambda_s & = & n_s \left\{ 1 -   \frac{1}{2}\frac{\beta e^2}{2 \pi^2 (\varepsilon_0+\varepsilon_1)}\left[k_{\rm max} - \DebFreq\tan^{-1}\frac{k_{\rm max}}{\DebFreq}\right] \right\},
\nonumber\\
\Lambda_d & = & n_d\left\{1  -  \frac{1}{4 \pi^2} \frac{\varepsilon_1}{n_d (\varepsilon_0+\varepsilon_1) }\left[ \frac{k_{\rm max}^3}{3}-k_{\rm max}(\DebFreq)^2 + \right.\right.
\nonumber\\
& & \left. \left.(\DebFreq)^3\tan^{-1}\frac{ k_{\rm max}}{\DebFreq}\right]\right\} \, .
\end{eqnarray}
The correction for the dielectric constant can be calculated by the same way as for the fugacity:
\begin{eqnarray}
\label{DeltaEpsOneLoopDef}
\varepsilon & = & \varepsilon_0 + \varepsilon_1 + \left.\frac{1}{2\beta} \int {\rm d}^3 \rbf_b \frac{\delta^2 \ln\left[\det(F^{(2)})\right]}{\delta E_i(\rbf_a) \delta E_i(\rbf_b)} \right|_{\phi_{\rm MF}=0},
\end{eqnarray}
The detailed calculation is presented in Appendix A and results in:
\begin{eqnarray}
\varepsilon  = \varepsilon_0 + \varepsilon_1 - \frac{3\beta \varepsilon_1^2}{2\Lambda_d} \nabla^2 g(0).
\end{eqnarray}
Substituting $\nabla^2 g(0)$ from Eq.~(\ref{Gvalue}), we get:
\begin{eqnarray}
\label{Delta_epsilon_Loop}
\varepsilon & = & \varepsilon_0+\varepsilon_1
\nonumber\\
& + & \frac{3  \DebFreq^3\varepsilon_1^2}{4 \pi^2 \Lambda_d (\varepsilon_0+\varepsilon_1)} \left[\frac{k_{\rm max}^3}{3\DebFreq^3} -\frac{k_{\rm max}}{\DebFreq} + \tan^{-1}\frac{k_{\rm max}}{\DebFreq}\right].
\end{eqnarray}
Adding the correction in the fugacity $\Lambda_d$, Eq.~(\ref{fugacity_one_loop}), to Eq.~(\ref{Delta_epsilon_Loop}) yields:
\begin{eqnarray}
\label{beyondPBeps_kmax}
\varepsilon & = & \varepsilon_0 + \varepsilon_1
\nonumber\\
& + & \frac{(\varepsilon_1)^2}{2 \pi^2 (\varepsilon_0+\varepsilon_1) n_d} \left[ \frac{k_{\rm max}^3}{3}- k_{\rm max}\DebFreq^2  +\DebFreq^3\tan^{-1}\frac{ k_{\rm max}}{\DebFreq}\right]\,. \nonumber \\
\end{eqnarray}
And finally, using the minimum cut-off length $a$, Eq.~(\ref{beyondPBeps_kmax}) yields~\cite{PRL}:
\begin{eqnarray}
\label{beyondPBeps}
\varepsilon & = & \varepsilon_0 + \varepsilon_1
\nonumber\\
& + & \frac{(\varepsilon_1)^2}{(\varepsilon_0+\varepsilon_1)} \frac{4\pi}{3 n_d a^3} \left[ 1 - \frac{3 }{4\pi^2}(a \DebFreq)^2   \right. \nonumber\\
 &+& \left. \frac{3}{8\pi^3}(a\DebFreq)^3\tan^{-1}\left(\frac{2\pi}{a \DebFreq}\right) \right]\, .
\nonumber\\
\end{eqnarray}

Equation~(\ref{beyondPBeps}) constitutes the principal result for the dielectric decrement as obtained using the one-loop expansion. The correction to the dielectric constant is composed of three terms. The first one represents the fluctuation effect of the water dipoles themselves beyond
the mean-field DPB level. It varies as $\sim 1/(n_d a^{3})$.
This pure water fluctuation term essentially adds a positive numerical prefactor of rather large magnitude to the mean-field value of $\varepsilon_0 +\varepsilon_1$  (of  about 12.1 for pure water). Hence,
it means that the one-loop correction is important even for the pure water case.

In the dilute salt limit, $\DebFreq a\ll 1$, we can further expand Eq.~(\ref{beyondPBeps})
to linear order in the salt concentration $n_s$, $\varepsilon(n_s)=\varepsilon_w+\gamma n_s$,
and get the coefficient $\gamma$ [as in   Eq.~(\ref{DecrementForm})]:
\begin{equation}
\label{beyondPBepsLinear}
\gamma=-\frac{\varepsilon_1^2 } {\varepsilon_0+\varepsilon_1} \frac{8\BejLen}{n_d a} \, ,
\end{equation}
where $\BejLen= \beta e^2/4\pi(\varepsilon_1+\varepsilon_0)$ is the Bjerrum length. The numerical value of $\gamma/\varepsilon_0$ is estimated to be -25\,M$^{-1}$, which is rather high, and indicates the importance of the additional non-linear term.
We treat $a$ as a free parameter and find its value by the best fit of our prediction,
while fixing the water dipolar moment to have its physical known value of $p_0=1.8$\,D.
The two additional correction terms in Eq.~(\ref{beyondPBeps}) account for water-ion correlations.
The leading term in the dilute
solution limit, $\DebFreq a\ll 1$, depends linearly on the salt concentration.
When the Debye length $\DebFreq^{-1}$ is of the
same order of magnitude as $a$, the last term in Eq.~(\ref{beyondPBeps})  starts to dominate and the
dielectric decrement becomes smaller until eventually it will reverse the trend and cause a dielectric
relative {\it increment}, as seen in some experiments~\cite{Experiment2} for high enough salt concentrations.

\section{Comparing One Loop Results to Experiments}
The static dielectric constant of an aqueous solution cannot be measured directly.
The effect of the static dielectric constant is measured by fitting high frequency data, and extracting the static dielectric constant as a fit parameter. Most experiments measure the dielectric response in microwave and RF frequencies, ranging from 100 MHz to 40 GHz~\cite{Experiment1, Experiment2, Experiment3}, with temperature in the range of $0^{\circ}$C - $60^{\circ}$C. The frequency-dependent permittivity is a complex function, which can be approximated by \cite{PolarMolecules}:
\begin{eqnarray}
\varepsilon(\omega) = \varepsilon_{\infty} + \frac{\varepsilon_s -
\varepsilon_{\infty}}{1+i\omega\tau} - i \frac{\sigma_{\mathrm{dc}}}{\varepsilon_0 \omega},
\end{eqnarray}
where $\varepsilon_{\infty}$ is the dielectric constant in the high frequency limit ($\omega \to \infty$), $\varepsilon_s=\varepsilon(\omega \to 0)$ is the static dielectric constant that is of interest to us, $\tau$ is the dielectric relaxation time, defined as the time that it takes for the dielectric response to reach equilibrium, $\sigma_{\mathrm{dc}}$ is the DC conductivity, and $\varepsilon_0$ is the vacuum permittivity.

The frequency-dependent permittivity can be measured and $\varepsilon_s$ can be obtained from a least-square fit. Table~\ref{tab:LiCl_and_RbCl} lists few such examples of the static $\varepsilon_s$ for  LiCl and RbCl salt solutions, in concentrations of $0.5$M, $1$M and $2$M.

\begin{table}[ht!]
\centering
\begin{tabular}{| p{1 cm}| l| p{1.5 cm} |}
\hline
$n_s$(M) & $\varepsilon_s$(LiCl) & $\varepsilon_s$(RbCl)  \\ \hline
0.5 & 71.2 & 73.5  \\ \hline
1 & 64.2 & 68.5 \\ \hline
2 & 51 & 58.5 \\ \hline
\end{tabular}
\caption{The static dielectric constant, $\varepsilon_s$, for aqueous salt solutions as fitted from $\varepsilon(\omega)$ measurements for different salts and concentrations\label{tab:LiCl_and_RbCl}. Adapted from Ref.~\cite{Experiment2}.}
\end{table}

\begin{figure*}[ht]
\includegraphics[scale=0.7,draft=false]{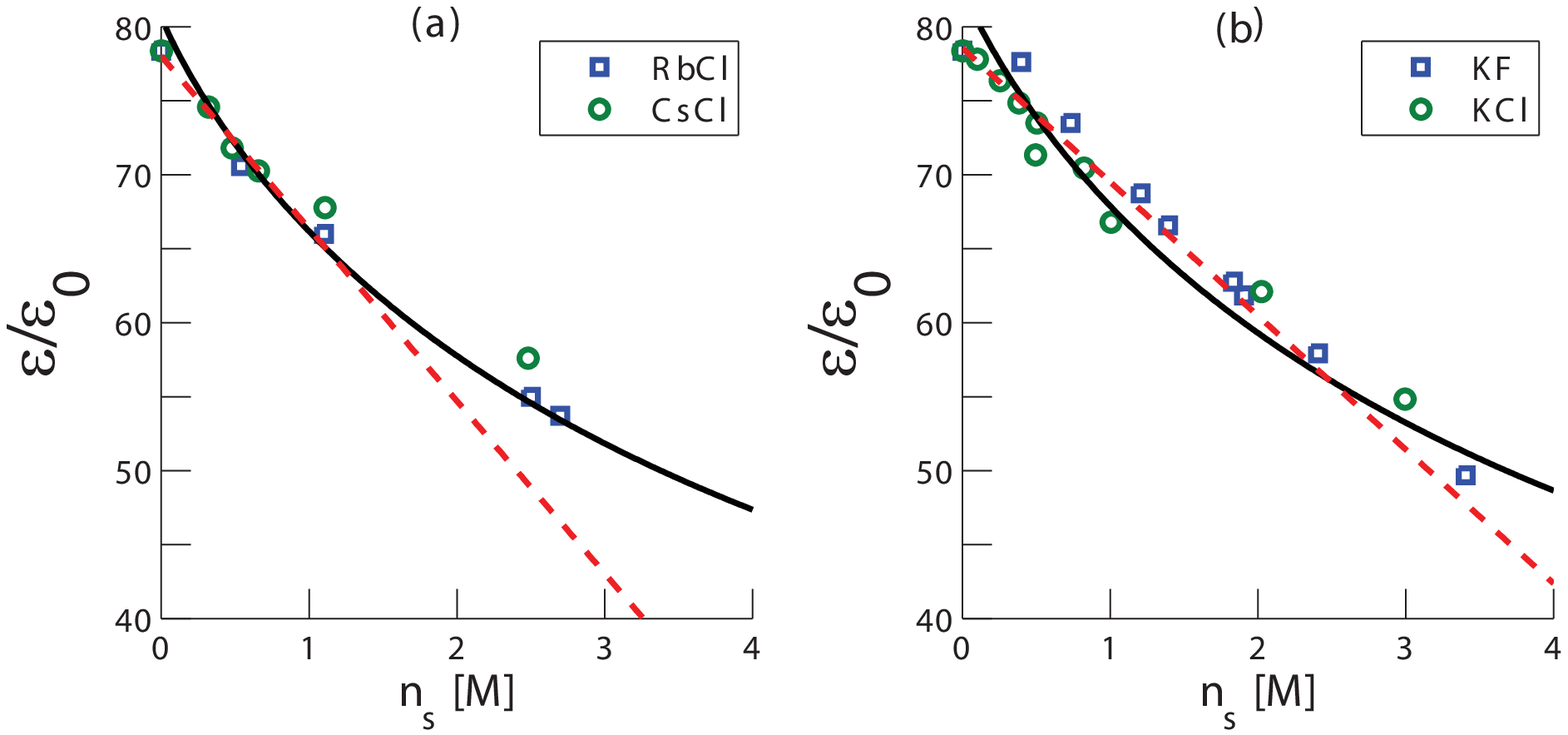}
\caption{\textsf{(color online) Comparison of the dielectric constant, $\varepsilon$, from the one-loop expansion, Eq.~(\ref{beyondPBeps}),
with experimental data for the static $\varepsilon_s$ from Ref.~\cite{Experiment2}, as function of ionic concentration, $n_s$, for various salts with larger ionic radii.
The theoretical prediction (solid line) was calculated
using the parameter $a$ as a fitting parameter. In (a) the best fit for RbCl and CsCl salts gives
$a=2.66$\,$\AArm$; while in (b) the best fit for KF and KCl gives $a=2.64$\,$\AArm$.
The dashed lines are the linear fit to the data in the low
$n_s\le 1$\,M range. The slope of the linear fit is $\gamma/\varepsilon_0=-11.7$\,M$^{-1}$ in (a) and $-9.0$\,M$^{-1}$ in (b).
The value of $\gamma$ for each salt
varies by about 10-20\,$\%$ and the linear fit should be taken as representative of the combined low $n_s$ behavior.
}}
\label{fig7}
\end{figure*}

\begin{figure*}[ht]
\includegraphics[scale=0.7,draft=false]{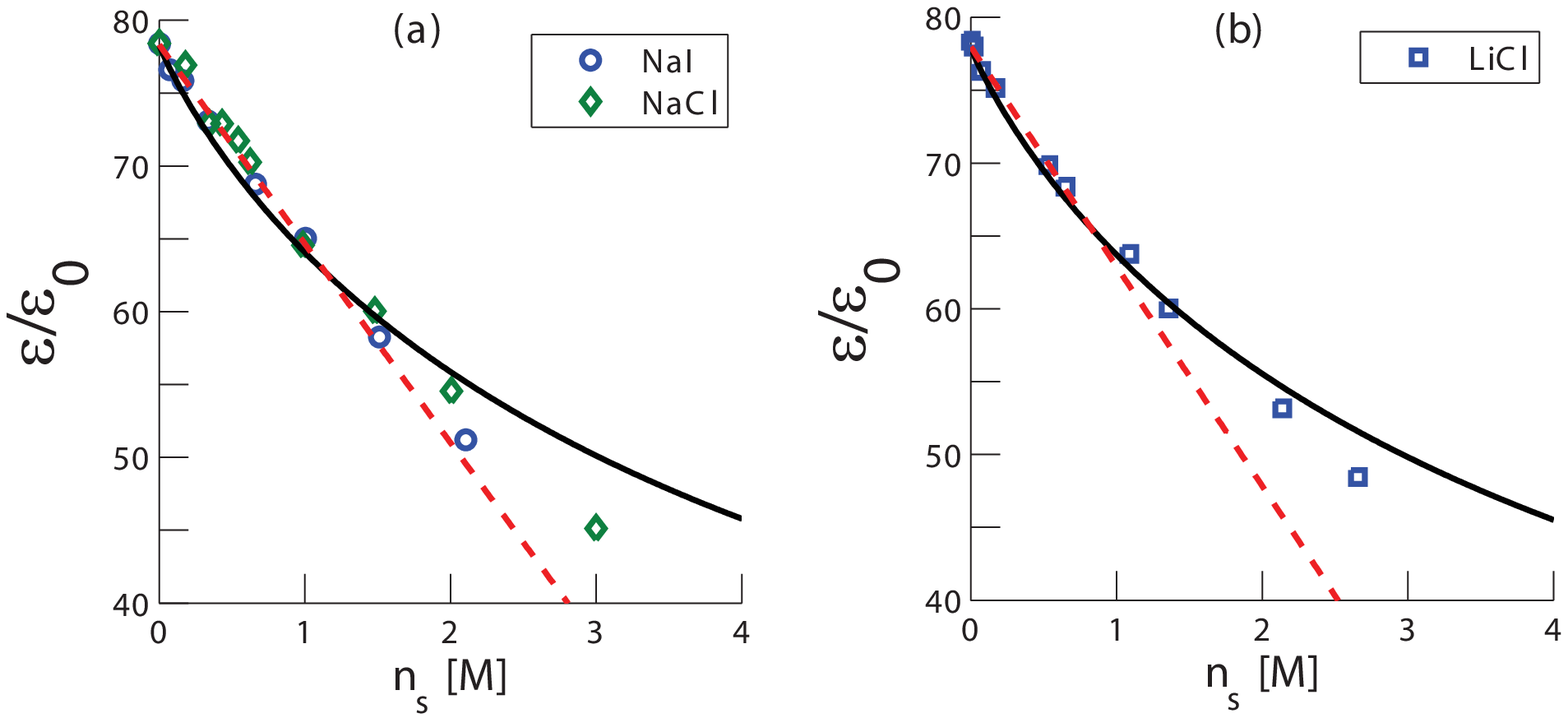}
\caption{\textsf{(color online) Comparison of the dielectric constant, $\varepsilon$, from the one-loop expansion, Eq.~(\ref{beyondPBeps}), with experimental data for the static $\varepsilon_s$ from Ref.~\cite{Experiment2}, as function of ionic concentration, $n_s$, for various salts with smaller ionic radii.
The theoretical prediction (solid line) was calculated
using the parameter $a$ as a fitting parameter. In (a) the best fit for NaI and NaCl gives $a=2.695$\,\AA; while in (b) the best fit for LiCl salt gives
$a=2.7$\,\AA. The dashed lines are the linear fit to the data in the low
$n_s\le 1$\,M range. The slope of the linear fit is $\gamma/\varepsilon_0=-13.65$\,M$^{-1}$ in (a) and $-15.1$\,M$^{-1}$ in (b).
}}
\label{fig8}
\end{figure*}

We compare our loop-expansion prediction for the dielectric constant $\varepsilon$, Eq.~(\ref{beyondPBeps}), to experimental values of the static $\varepsilon_s$~\cite{Experiment2}
for seven different ionic solutions in a concentration range of 0--4\,M.
We separate the seven salts into four subgroup according to the
size of the alkaline cations, and present the results in Figs.~\ref{fig7} and ~\ref{fig8}. In each of the figure parts the parameter $a$ is fitted separately. We treat $a$
as a free parameter and find its value by the best fit of our prediction,  Eq.~(\ref{beyondPBeps}), to experimental data, while keeping the physical known value of the water dipolar moment, $p_0=1.8$\,D.

The largest ionic size of Cs$^+$ and Rb$^+$ gives the best results [Fig.~\ref{fig7}(a)], and the fit remains good even for high concentrations of about 3-4\,M. In Fig.~\ref{fig7}(b) the fit for K$^+$ ions (for two solutions with anions F$^-$ and Cl$^-$) is also quite good, although some deviations are seen, especially in the dilute limit. We also show for comparison a linear fit to the data. Note that this linear fit is done without any modeling or external parameter. It is not the same as the linearized term obtained from our model, Eq.~(\ref{beyondPBepsLinear}). The  latter gives $\gamma/\varepsilon_0\simeq -25$\,M$^{-1}$, and does not fit the data as well.
In Fig.~\ref{fig8}, the fit for the two smaller cations Li$^+$ and Na$^+$ (for LiCl, NaCl and NaI solutions) works well only up to $n_s=2$\,M, but for higher $n_s$ the fit over-estimates the experimental $\varepsilon$. We also get a good fit for pure water $\varepsilon_w \simeq 78$, which is an important result since we are using only one fitting parameter, $a\simeq 2.6-2.7$\AA. For the K$^+$ case, the fit for the low salt limit does not fit so well and for pure water the best fit overestimates the water value, $\varepsilon_w\simeq 83$.

The ionic size effect can be understood from a microscopic point of view. As the field in the vicinity of the ion is high, an approximate calculation, such as the one-loop expansion, is more likely to fail. It can be related to the significant deviation we have seen in comparing numerical and approximate analytical solution of the DPB equation (Fig.~\ref{fig2}).
Moreover, note that our formula takes into account only in a broad sense the finite size of ions (and the distance of closest approach between them)
via a single parameter, $a$, which effectively combines the dipole and ion sizes.
It is beyond the level of the theory to give more specific ionic predictions.
Hence, the obtained value of  $a\simeq 2.7$\,\AA\  is not very sensitive to the type of salt. Rather, its main contribution comes from the water dipoles themselves whose
diameter is about $2.75$\,\AA~\cite{WaterRadii}.
On the other hand, as can be clearly seen from Figs.~\ref{fig7} and~\ref{fig8}, important cooperative effects
of ions and dipoles are accounted for in our non-linear expression for $\varepsilon(n_s)$.
For small $n_s$, the dashed line represents the best linear fit and works well only when $n_s\le 1$\,M, while the non-linear prediction (solid line) of
Eq.~(\ref{beyondPBeps}) succeeds in fitting the large concentration range as well.

\section{Conclusions}

The decrement of the dielectric constant in ionic solutions is a well-observed phenomena, studied both theoretically and experimentally. Since the pioneering works of Debye, Onsager and Kirkwood~\cite{PolarMolecules, Onsager, Kirkwood}, to more recent works using molecular dynamics (MD)~\cite{MD1,MD2, MD3}, different approaches were advanced to explain this effect. In this paper, we addressed the dielectric constant of an ionic solution from a field-theory point of view. Starting from a general system composed of different types of charges and dipoles interacting via electrostatic Coulomb interactions, we modeled the ionic solution as a system of charged particles surrounded by dipoles. After writing the grand-canonical partition function as a functional integral on the electrostatic potential, it was possible to extract physical quantities on the mean-field level and also to find corrections that go beyond mean-field and include correlations and fluctuations on the one-loop level. Furthermore, we investigated how these different effects give rise to variations in the dielectric constant of different ionic solutions.

On a mean-field level, the key feature of our model is that it accounts for any internal charge distribution of particles, rather than only point-like or rod-like particles~\cite{rodlike}. A generalized PB equation is derived, and serves as a convenient starting point for our discussion of ionic solutions.  The DPB equation is a special case of the generalized PB equation that is explored in great detail.
By looking at the DPB equation around a point-like ion at the origin, a closed-form formula for the dielectric constant is obtained. We expressed the dielectric constant using several physical length scales. The most important one is the ``hydration length" $l_h$, which characterizes the hydration shell of dipoles around ions, and thus the strength of the dielectric decrement. From the DPB equation the dielectric response is then calculated for three additional cases: mixture of polar solvents, polarizable medium and ions of finite size.

Beyond mean-field theory, using loop-expansion analysis, we are able to  derive analytically the dielectric constant. The expression for the dielectric constant is found to be in good agreement with the experimental data, in a wide range of ionic concentrations. However, specific behavior of different salts, which can be accounted for in other frameworks such as MD simulations, cannot be predicted by our model.

Correlations are evidently a key mechanism in understanding the electrostatic behavior of ionic solutions, and the loop-expansion technique of field theory is a useful tool  for investigating them. Removing some of the underlying limitations of our theory  may reveal more interesting phenomena. One of the  model limitations  is that only  first-order corrections to mean-field theory were considered. Taking additional terms beyond the one-loop expansion might be useful to access the validity of the approximation. However, as water molecules are modeled as point-like dipoles, the neglect of the finite size of the water dipoles might be of greater importance than higher-order loop corrections.

Another remark on the one-loop expansion is that it has a single free parameter, the cutoff distance $a$, which was added in order to avoid the divergence of the integrals. A more elegant way of regulating the divergence is to consider explicitly the self energy in the partition function~\cite{SelfEnergy}.

A further interesting application of our model is to examine the dielectric constant near a charged surface. We restrict ourselves only to bulk properties, where we could extract analytical solutions. However, interesting physical processes occur near charged membranes of biological cells, and the extrapolation from the bulk is far from being straightforward~\cite{IonSpecific}.

Finally, we propose possible extensions to include ion-specific effects. We have started with a generalized model of ionic solutions that allows any kind of charge distribution, while focusing only on ionic solutions composed of point-like or sphere-like particles. This assumption did not allow for major ionic specific effects. Hence, it may be of interest to expand the finite-size effects to the one-loop approximation as well. Another venue of interest may be to include additional non-Coulombic interactions that can lead to significant corrections and interesting modifications, going beyond the scope of the present work.

\bigskip
{\bf Acknowledgements.~~~} {We thank D. Ben-Yaakov, Y. Burak, X. K. Man and R. Podgornik for useful discussions. This work
was supported in part by the Israel Science Foundation under Grants No. 231/08 and 438/12. One of us (HO) would like to thank  the Raymond \& Beverly Sackler Program for Senior Professors by Special Appointment at Tel Aviv University.}

\appendix

\section{Dielectric constant correction }

Calculating  the correction of the dielectric constant is less straight forward than that of the fugacity, and will be explained in detail in this appendix. The correction term about the MFT result, $\varepsilon_{\rm MF}=\varepsilon_0+\varepsilon_1$ , was given by Eq.~(\ref{DeltaEpsOneLoopDef}):
\begin{eqnarray}
\Delta \varepsilon & = & \varepsilon - \varepsilon_{\rm MF} =  \left. \frac{1}{2\beta}\int{\rm d}^3 \rbf_b  \frac{\delta^2 \ln\left[\det(F^{(2)})\right]}{\delta E_i(\rbf_a) \delta E_i(\rbf_b)} \right|_{\phi=\phi_{\rm MF}} \,. \nonumber\\
\end{eqnarray}

The second functional derivative $F^{(2)}$, Eq.~(\ref{DPBfunctinal2}), can be rewritten using the electrostatic potential $\psi$, and the electric field $\mathbf{E}$:
\begin{eqnarray}
F^{(2)}(\rbf,\rbf ')  & = &  -\varepsilon_0 \nabla^2\delta( \mathbf{r-r'})
\nonumber\\
& + &    2\Lambda_s \beta e^2\cosh \left[ \beta e \psi(\rbf)\right] \delta( \mathbf{r-r'})
\nonumber\\
& + &  \Lambda_d \beta \int {\rm d}^3 \mathbf{r''} \int \frac{{\rm d}^2\, \Omega}{4\pi}\, \left[\mathbf{p}_0 \cdot \nabla  \delta( \mathbf{r-r''}) \right]
\nonumber\\
& \times &{\rm e}^{-\beta {\bf p}_0 \cdot \mathbf{E}} \left[\mathbf{p}_0 \cdot \nabla  \delta( \mathbf{r'-r''})\right]\,.
\nonumber\\
\end{eqnarray}
As the field $\mathbf{E}$ appears only in 3$rd$ term of $F^{(2)}$, the correction to the dielectric constant will be derived from it:
\begin{eqnarray}
 \Delta \varepsilon & =& -\frac{\Lambda_d \beta}{2}\int {\rm d}^3\rbf_b \nonumber\\
 & & \times \frac{\delta^2}{\delta E_i(\rbf_a) \delta E_i(\rbf_b)} \int {\rm d}^3 \rbf \, \int {\rm d}^3 \mathbf{r'}\, g(\mathbf{r,r'})   \int {\rm d}^3 \mathbf{r''}
\nonumber\\
& & \times\int {\rm d}^2 \Omega \,\left[\mathbf{p}_0 \cdot \nabla  \delta( \mathbf{r-r''}) \right]{\rm e}^{- \beta {\bf p}_0 \cdot {\bf E}} \left[\mathbf{p}_0 \cdot \nabla  \delta( \mathbf{r'-r''})\right]
\nonumber\\
& = &-\frac{\Lambda_d \beta^3}{2}\int \frac{{\rm d}^2 \Omega}{4\pi} \, p_{0i}^2  \int {\rm d}^3 \rbf \int {\rm d}^3 \mathbf{r'} \int {\rm d}^3 \mathbf{r''}
\nonumber\\
& & \times
\left[\mathbf{p}_0 \cdot \nabla  \delta( \rbf'-\rbf'')\right]  g(\mathbf{r,r'})   \left[ \mathbf{p}_0 \cdot \nabla  \delta( \mathbf{r-r''}) \right] \delta(\rbf-\rbf_a) \, .
\nonumber\\
\end{eqnarray}
Substituting $\phi=\phi_{\rm MF} =0$, and using integration by parts we get:
\begin{eqnarray}
\label{DeltaEpsAppx}
\Delta \varepsilon & = &\frac{\Lambda_d \beta^3}{2 }  \int {\rm d}^3 \rbf \int \frac{{\rm d}^2 \Omega}{4\pi} \,
\nonumber\\
& & \times
p_{0i}^2\left[\left(\mathbf{p}_0 \cdot \nabla \right)^2 \delta( \rbf-\rbf_a)\right]  g(\rbf,\rbf_a)\,.
\end{eqnarray}
Defining
\begin{equation}
\label{IijDef}
I_{i}  = -   \int \frac{{\rm d}^2 \Omega}{4\pi} \,\, p_{0i}^2 (\mathbf{p}_0 \cdot \nabla )^2 \delta( \rbf-\rbf_a )\, ,
\end{equation}
where $i=x,y,z$, and for isotropic systems we can restrict the treatment to $i=z$.
Substituting $\delta(\rbf-\rbf_a) = \int \frac{{\rm d}^3 \mathbf{k}}{(2\pi)^3}\, {\rm e}^{i \mathbf{k} \cdot (\rbf-\rbf_a)}$ in (\ref{IijDef}), yields:
\begin{eqnarray}
I_{z} & = &
\nonumber\\
& - &  \int \frac{{\rm d}^2 \Omega}{4\pi} \,\, p_{0z}^2 (\mathbf{p}_0 \cdot \nabla )^2  \int \frac{{\rm d}^3 \mathbf{k}}{(2\pi)^3} \, {\rm e}^{i \mathbf{k} \cdot (\rbf-\rbf_a)}
\nonumber\\
& = &    \int \frac{{\rm d}^3 \mathbf{k}}{(2\pi)^3} \int \frac{{\rm d}^2 \Omega}{4\pi} \,\, p_{0z}^2 (\mathbf{p}_0 \cdot \mathbf{k})^2  {\rm e}^{i \mathbf{k} \cdot (\rbf-\rbf_a)}.
\nonumber\\
\end{eqnarray}
We choose $\hat{z}$ direction to be in the direction of $\mathbf{r-r'}$, and the scalar product between ${\bf p}$ and ${\bf k}$ depends on two sets of polar angles: the polar angles of $\mathbf{p}$ defined as $\theta$ and $\varphi$, and the ones of  $\mathbf{k}$ defined as $\alpha$ and $\beta$:
\begin{eqnarray}
\mathbf{p \cdot k} & = &  pk \left[\sin\theta \sin\alpha \cos(\beta- \varphi) + \cos\theta\cos\alpha\right].
\end{eqnarray}
The integral then becomes:
\begin{eqnarray}
I_{z} & = & \frac{1}{32\pi^4} \int k^2 {\rm d}k\, {\rm d} (\cos\alpha) \, {\rm d}\beta\, \int {\rm d} (\cos\theta) \, {\rm d}\varphi \, p_0^4 k^2  \cos^2\theta
\nonumber\\
&\times & \left[\sin^2\theta \sin^2\alpha \cos^2(\beta- \varphi) + \cos^2\theta\cos^2\alpha \right.
\nonumber\\
& + & \left. \frac{1}{2} \sin2\theta \sin2\alpha \cos(\beta- \varphi) \right]{\rm e}^{i\mathbf{k} \cdot (\rbf-\rbf_a)}.
\end{eqnarray}
First, we can integrate over $\varphi$. The integration of the first term, $\cos^2(\beta - \varphi)$, is equal to $\pi$. The integration of the second term does not depend on $\varphi$ and equals to $2 \pi$, while integrating the third term gives zero.
\begin{eqnarray}
I_{z} & = & \frac{1}{4(2\pi)^3} \int k^2 {\rm d} k\, {\rm d} (\cos\alpha)\, {\rm d} \beta\, \int {\rm d} (\cos\theta) \, p_0^4 k^2  \cos^2\theta
\nonumber\\
& \times & \left[\sin^2\theta \sin^2\alpha + 2 \cos^2\theta\cos^2\alpha\right] {\rm e}^{i\mathbf{k} \cdot (\rbf-\rbf_a)}.
\end{eqnarray}
Next, the integration of $\cos^4\theta$ gives $8\pi/5$ and the integration over $\sin^2\theta\cos^2\theta$ is equal to $16\pi/15$:
\begin{eqnarray}
I_{z} & = & \frac{1}{(2\pi)^3}\int k^2 {\rm d}k\, {\rm d} (\cos\alpha)\, {\rm d}\beta \,
\nonumber\\
&\times& \int  p_0^4 k^2  \left[\frac{1}{15} \sin^2\alpha  +\frac{1}{5} \cos^2\alpha \right] {\rm e}^{i\mathbf{k} \cdot (\rbf-\rbf_a)}.
\end{eqnarray}
We can rearrange the terms such that one term depends only on $k$, and another depends only on $k \cos\alpha$:
\begin{eqnarray}
I_{z} & = & \frac{1}{5} \int \frac{{\rm d}^3 \mathbf{k}}{(2\pi)^3}\,  p_0^4 k^2  {\rm e}^ {i\mathbf{k} \cdot (\rbf-\rbf_a)}
\nonumber\\
& - & \frac{2}{15}\int \frac{{\rm d}^3 \mathbf{k}}{(2\pi)^3} \, p_0^4 (k \cos\alpha)^2  {\rm e}^{i\mathbf{k} \cdot (\rbf-\rbf_a)}.
\end{eqnarray}
The first term is associated with the Laplacian of the $\delta$-function, and the second one with the second derivative of the $\delta$-function with respect to $r$:
\begin{eqnarray}
\label{A12}
I_{z} & = & -\frac{p_0^4}{5} \left[ \nabla^2 \delta(\rbf-\rbf_a) + \frac{2 }{3}\partial^2_{r}\delta(\rbf-\rbf_a) \right].
\end{eqnarray}
We can now substitute Eq.~(\ref{A12}) into the correction of the dielectric constant Eq.~(\ref{DeltaEpsAppx}), and get:
\begin{eqnarray}
& & \Delta \varepsilon  =  -\frac{\Lambda_d \beta^3 p_0^4}{10}\int {\rm d}^3 \rbf \,g(\rbf,\rbf_a)
\nonumber\\
& \times &  \left[ \nabla^2 \delta(\rbf-\rbf_a) + \frac{2 }{3}\partial^2_{r}\delta(\rbf-\rbf_a) \right]
\nonumber\\
& = & - \frac{\Lambda_d \beta^3 p_0^4}{10} \int {\rm d}^3 \rbf \, g(\rbf,\rbf_a)
\nonumber\\
& \times & \left[ \nabla^2 \delta(\rbf-\rbf_a) + \frac{2 }{3}\partial^2_{r}\delta(\rbf-\rbf_a) \right]
\nonumber\\
& = & - \frac{\beta^3 p_0^4}{10} \Lambda_d
\nonumber\\
&\times & \int {\rm d}^3 \rbf \, g(\mathbf{r,r}_a) \left( \nabla^2  + \frac{2 }{3}\partial^2_{r} \right)\delta(\mathbf{r-r}_a) .
\end{eqnarray}
Using integration by parts twice, and the fact that $g(r)$ depends only on $r$ so that $\nabla^2 g(0) = d^2g(0)/dr^2$, we finally get:
\begin{eqnarray}
\Delta \varepsilon & = & - \frac{\Lambda_d\beta^3 p_0^4}{6} \nabla^2 g(0)
\nonumber\\
& = & - \frac{3 \beta\varepsilon_1^2 }{2\Lambda_d} \nabla^2 g(0).
\end{eqnarray}
%


\end{document}